\documentstyle[preprint,aps,prd,psfig]{revtex}
\newcommand{\la}{\langle}
\newcommand{\ra}{\rangle}

\newcommand{\AmS}{{\protect\the\textfont2
  A\kern-.1667em\lower.5ex\hbox{M}\kern-.125emS}}
\newcommand{\be}{\begin{equation}}
\newcommand{\ee}{\end{equation}}
\newcommand{\ben}{\begin{eqnarray}}
\newcommand{\een}{\end{eqnarray}}
\newcommand{\nn}{\nonumber}

\newcommand{\slas}[2]{{{#1}\hspace{-5pt}{/}}_{#2}}
\newcommand{\slal}[2]{{{#1}\hspace{-8pt}{/}}_{#2}}
\def\simgt{\rlap{\lower 3.5 pt\hbox{$\mathchar \sim$}}\raise 1pt \hbox {$>$}}
\def\simlt{\rlap{\lower 3.5 pt\hbox{$\mathchar \sim$}}\raise 1pt \hbox {$<$}}

\newcommand{\mm}{{\cal M}}
\newcommand{\ce}{{\it c}_{\it E}}
\newcommand{\cb}{{\it c}_{\it B}}
\newcommand{\lqcd}{{\Lambda}_{\rm QCD}}

\begin{document}
\draft
 
\title{
\vspace{-5.5cm}
%\begin{flushright}
\hspace{120mm}{\normalsize KEK-CP-110}\\
\vspace{+8mm}
%\end{flushright}
Relativistic Heavy Quarks on the Lattice}

\author{Sinya~Aoki$^{\rm a}$,
        Yoshinobu~Kuramashi$^{\rm b}$ and
        Shin-ichi~Tominaga$^{\rm c}$
       }
\address{$^a$Institute of Physics,
             University of Tsukuba,\\
             Tsukuba, Ibaraki 305-8571, Japan\\
         $^b$Institute of Particle and Nuclear Studies, 
             High Energy Accelerator Research Organization(KEK),\\ 
             Tsukuba, Ibaraki 305-0801, Japan\\
         $^c$Center for Computational Physics,
             University of Tsukuba,\\
             Tsukuba, Ibaraki 305-8577, Japan\\
        }

\date{\today}

\maketitle

\vspace{-7mm}

%\newpage
\begin{abstract}
\vspace{-10mm}

Lattice QCD should allow quantitative predictions for the
heavy quark physics from first principles. 
Up to now, however, most approaches have based on the
nonrelativistic effective theory, with which
the continuum limit can not be taken in principle. 
In this paper we investigate feasibility of relativistic approaches to
the heavy quark physics in lattice QCD.
We first examine validity of the idea that
the use of the anisotropic
lattice could be advantageous to control the $m_Q a$ corrections.
Our perturbative calculation, however, reveals that this is not true.
We instead propose a new relativistic approach to handle
heavy quarks on the isotropic lattice.
We explain how power corrections of $m_Q a$ can be avoided and 
remaining uncertainties are reduced to be of order 
$(a\Lambda_{\rm QCD})^2$.

\end{abstract}

\section{Introduction}

Weak matrix elements associated with $B$ mesons are essential ingredients
to determine Cabibbo-Kobayashi-Maskawa matrix
elements. In principle lattice QCD provides the opportunity
of evaluating these matrix elements from first principles.   
However it is still difficult to simulate the $b$ quark with
high precision on the lattice. The main source of systematic errors
originates from the fact that the $b$ quark mass in the
lattice unit is large: $m_b a\sim 1-2 $ in the quenched
approximation and $m_b a\sim 2-3 $ in full QCD with current
accessible computational resources. In order to control 
large $m_Q a$ errors, several ways have been proposed so far:
A static approximation with $m_Q\rightarrow \infty$\cite{static}, 
a nonrelativistic QCD\cite{nrqcd}, a nonrelativistic interpretation
applied to results with the
Wilson/Sheikholeslami-Wholert (SW) action\cite{fermilab} and an anisotropic
lattice with finer temporal lattice spacing $a_t$ while keeping
the spatial one $a_s$ modest\cite{aniso}.
Although the $b$ quark can be directly  simulated with any of the last
three approaches, only the last one has the advantage that we can take the
continuum limit, which is a fascinating feature stimulating our interest. 

Practical effectiveness of the anisotropic lattice is transparent:
with finer temporal lattice spacing   
time evolutions of all kinds of correlation
functions become milder, which benefits us the better
signal-to-noise ratio.
On the theoretical side, our interest exists in the use of the
anisotropic lattice to control $m_Q a$ errors. If they
are restricted to only powers of $m_Q a_t$,
they can be made small by the anisotropic lattice with smaller $a_t$.
Indeed
$m_Q a_s$ corrections can be removed at the tree level\cite{symm_at}. 
Our main concern, however, is whether
$m_Q a_s$ corrections could revive perturbatively or nonperturbatively.
Up to now no one has successfully
eliminated the possibility that $m_Q a_s$
corrections can appear beyond the tree level.
Without the proof given the anisotropic lattice is no better
than the isotropic one, where one can also
eliminate $m_Q a$ corrections at the tree level, 
in terms of controlling $m_Q a$ corrections.

The first part of this paper is devoted to a one-loop
calculation of the quark self energy on the anisotropic
lattice. We analyze its $O(g^2 a)$ terms to examine
the possibility of appearance of  $O(g^2 m_Q a_s)$
contributions. 
Our results strongly suggest that      
one-loop radiative corrections allow the revival of 
$m_Q a_s$ corrections.
Since we find that the anisotropic lattice is not theoretically advantageous 
any more, 
in the second part of this paper
we propose a new relativistic way to deal with the heavy quarks on the 
isotropic lattice,
analyzing $m_Q a$ corrections carefully.
We discuss cutoff  effects of the heavy quark system
following the on-shell improvement programme\cite{sym1,sym2}.
An important finding is that leading cutoff effects of
order $(m_Q a)^n$ can be absorbed
in the definition of renormalization factors for the
quark wave function and mass, so that the leading $m_Q a$ correction
is reduced to be of order $ (m_Q a)^n a \Lambda_{\rm QCD} $.
After removing remaining leading cutoff effects of
$O((m_Q a)^n a\Lambda_{\rm QCD})$ with parameters in the
quark action properly adjusted, 
we are left with 
only $O((a\Lambda_{\rm QCD})^2)$ errors, which are
expected to be fairly small.
We also provide a nonperturbative method to control the $O((m_Q a)^n)$
corrections involved in renormalization factors.

This paper is organized as follows.
In Sec.~\ref{sec:aniso} 
we present a one-loop calculation of the quark self energy
on the anisotropic lattice
and discuss the possibility of the
revival of $m_Q a_s$ corrections.
In Sec.~\ref{sec:iso} we propose a new relativistic
approach to handle the heavy quarks on the isotropic lattice
avoiding large $m_Q a$ corrections.   
Our conclusions are summarized in Sec.~\ref{sec:conclusion}.

\section{Anisotropic lattice}
\label{sec:aniso}

\subsection{On-shell improvement on the anisotropic lattice}
\label{subsec:onshell}

In order to obtain a generic form of the quark action
allowed on the anisotropic lattice, let us make the
operator analysis according to the Symanzik's 
improvement programme\cite{sym1,sym2}. 
The lattice theory is described by a local
effective theory as
\be
S_{\rm eff}=S_0+\sum_{k\ge 1,i} a^k \int d^4 x c_{4+k,i}(g){\cal O}_{4+k,i} (x),
\ee
where $S_0$ denotes the continuum action. 
${\cal O}_{k,i} (x)$ is a  
local composite operator with $k$ dimensions, which consist of
quark mass, quark fields and link variables.
These higher dimensional operators must respect 
symmetries on the lattice such as the invariance under
gauge, parity and charge-conjugation transformations and
discrete rotations.
The coefficient $c_{k,i}$ 
is a function of the gauge coupling $g$
to be determined perturbatively or nonperturbatively.   

Symanzik's improvement programme was originally
designed to reduce cutoff effects order by order for on-shell and
off-shell Green functions. 
However, we are interested in 
only on-shell quantities such as hadron masses and matrix elements
which require correlation functions at non-zero
physical distances.
Here it would be better to consider the on-shell improvement
procedure that is much simpler but restricted to
on-shell quantities\cite{onshell}.

Under the requirement of various symmetries on the lattice,
we find the following set of operators with dimension up to
five:
\ben
{\rm dim.3:}&& {\cal O}_3^\prime(x)={\bar q}(x)q(x), \\ 
{\rm dim.4:}&& {\cal O}_{4a}^\prime(x)={\bar q}(x)\gamma_0 D_0 q(x),\\
            && {\cal O}_{4b}^\prime(x)=
                 \sum_i {\bar q}(x)\gamma_i D_iq(x), \\ 
{\rm dim.5:}&& {\cal O}_{5a}^\prime(x)={\bar q}(x)D_0^2 q(x),\\
            && {\cal O}_{5b}^\prime(x)=\sum_i {\bar q}(x) D_i^2 q(x), \\
            && {\cal O}_{5c}^\prime(x)=
                 i\sum_i {\bar q}(x)\sigma_{0i}F_{0i} q(x),\\
            && {\cal O}_{5d}^\prime(x)=
                 i\sum_{i,j} {\bar q}(x)\sigma_{ij}F_{ij} q(x),\\
            && {\cal O}_{5e}^\prime(x)=
                 \sum_i {\bar q}(x) [\gamma_0 D_0,\gamma_i D_i] q(x), 
\een  
where $D_\mu$ is the covariant derivative on the lattice and 
$\sigma_{\mu\nu}$ and $F_{\mu\nu}$ are defined as
$\sigma_{\mu\nu}\equiv[\gamma_\mu,\gamma_\nu]/2$ 
and $ig F_{\mu\nu}\equiv[D_\mu,D_\nu]$.
The subscript $0$ denotes the time component, while $i,j=1,2,3$
space components.
In terms of these operators, a general form of the quark
action on the anisotropic lattice is given by
\ben
S_q^{\rm imp}&=&\sum_x\left[ 
c_3^\prime{\cal O}_3^\prime(x)
+\sum_{i=a,b}c_{4i}^\prime{\cal O}_{4i}^\prime(x) 
+\sum_{i=a,\dots ,e}c_{5i}^\prime{\cal O}_{5i}^\prime(x) 
\right],
\label{eq:qaction_g}
\een
where $c_3^\prime,\dots,c_{5e}^\prime$ are functions of the bare gauge
coupling $g$, the bare quark mass $m_0$ and the time and
space lattice spacings $a_t$, $a_s$.
Since degrees of freedom of $c_3^\prime$ and $c_{4a}^\prime$ can be 
absorbed in the renormalization of the quark mass and the
wave function respectively, we choose $c_3^\prime=m_0$ and 
$c_{4a}^\prime=1$ for
convenience. We also find ${\cal O}_{5a}^\prime$ and 
${\cal O}_{5e}^\prime$ are related to
other operators by using the classical field
equation, and hence $c_{5a}^\prime$ and  $c_{5e}^\prime$ can be
set by hand. We eliminate ${\cal O}_{5e}^\prime$ by choosing 
$c_{5e}^\prime=0$ and employ ${\cal O}_{5a}^\prime$ to avoid
species doubling with  $c_{5a}^\prime$ finite. 
After all remaining parameters are $c_{4b}^\prime$,
$c_{5b}^\prime$, $c_{5c}^\prime$ and $c_{5d}^\prime$, which should be    
tuned (nonperturbatively) in order to remove 
$O(a_{t,s})$ discretization errors,
so that we are left with only discretization ambiguities of $O(a_{t,s}^2)$.
This point should be stressed, since previous
papers\cite{symm_at,pp} claim that only three parameters 
$c_{4b}^\prime$, $c_{5c}^\prime$ and $c_{5d}^\prime$ are enough to be tuned
for the $O(a_{t,s})$ improvement.

\subsection{Quark and gauge actions}

According to the discussion in the above subsection
a general form of the
quark action on the anisotropic lattice is given by:
\ben
S_q&=& a_t a_s^3 \sum_x {\bar q}(x)\left[ \gamma_0 D_0
+\nu \sum_i \gamma_i D_i +m_0 
-\frac{a_t}{2}r ( D_0^2+\eta \sum_i D_i^2)\right.\nn\\
&&\;\;\;\;\left.-a_t \frac{ig}{4}r 
\left\{ \ce(1+\eta)\sum_i \sigma_{0i} F_{0i}
+\cb \eta \sum_{ij} \sigma_{ij} F_{ij}
\right\}\right]q(x),
\label{eq:qaction_d}
\een  
where parameters in eq.~(\ref{eq:qaction_g}) are rewritten as
\ben
c_{4b}^\prime&=& \nu,\\
c_{5a}^\prime&=& -\frac{r}{2},\\ 
c_{5b}^\prime&=& -\frac{r \eta}{2},\\
c_{5c}^\prime&=& -\frac{g}{4} r \ce (1+\eta),\\
c_{5d}^\prime&=& -\frac{g}{4} r \cb \eta.
\een
Here the Wilson parameter $r$ is taken arbitrary by 
hand as mentioned in the previous subsection.
Note that the quark fields $q$,  the covariant derivatives
$D_0$ and $D_i$, the bare quark mass $m_0$, the gauge field
strengths $F_{0i}$ and $F_{ij}$ are dimensionful.  
They are transformed into
dimensionless quantities by
\ben
{\tilde q}&=&a_s^{3/2} q,\\
{\tilde D}_0&=& a_t D_0,\\
{\tilde D}_i&=& a_s D_i,\\
{\tilde m}_0&=& a_t m_0,\\
{\tilde F}_{0i}&=& a_t a_s F_{0i},\\
{\tilde F}_{ij}&=& a_s^2 F_{ij}.
\een  
With this transformation, eq.~(\ref{eq:qaction_d}) is
rewritten as 
\ben
S_q(x)&=& \sum_x {\bar {\tilde q}}(x)\left[ \gamma_0 {\tilde D}_0
+\frac{\nu}{\xi}\sum_i \gamma_i {\tilde D}_i +{\tilde m}_0 
-\frac{r}{2}( {\tilde D}_0^2
+\frac{\eta}{\xi^2} \sum_i {\tilde D}_i^2)\right.\nn\\
&&\;\;\;\;\left.- \frac{ig}{4}r 
\left\{ \ce \frac{(1+\eta)}{\xi}\sum_i \sigma_{0i} {\tilde F}_{0i}
+\cb \frac{\eta}{\xi^2} 
\sum_{ij} \sigma_{ij} {\tilde F}_{ij} \right\}\right]
{\tilde q}(x)
\label{eq:qaction}
\een  
with $\xi=a_s/a_t$ the anisotropy parameter.

 From eq.~(\ref{eq:qaction_d})
we find
the inverse free quark propagator
\ben
a_t S_q^{-1}(p)&=&i \gamma_0 {\rm sin}(p_0 a_t)
+\frac{\nu}{\xi} 
i \sum_i \gamma_i {\rm sin}(p_i a_s) +m_0 a_t\nn\\
&&+r (1-{\rm cos}(p_0 a_t))+\frac{r\eta}{\xi^2}
\sum_i(1-{\rm cos}(p_i a_s)),
\label{eq:qp_tree}
\een
and relevant vertices for the present calculation 
\ben
V_{10}^{A}(q,p)&=&-g T^A\left\{i \gamma_0 
{\rm cos}\left(\frac{p_0 a_t + q_0 a_t}{2}\right)
+r {\rm sin}\left(\frac{p_0 a_t + q_0 a_t}{2}\right)\right\},\\
V_{1i}^{A}(q,p)&=&-g T^A\left\{\nu i \gamma_i 
{\rm cos}\left(\frac{p_i a_s + q_i a_s}{2}\right)
+\frac{r\eta}{\xi} 
{\rm sin}\left(\frac{p_i a_s + q_i a_s}{2}\right)\right\},\\
V_{200}^{AB}(q,p)&=&\frac{a_s}{2\xi} g^2 \frac{1}{2}
\left\{T^A,T^B\right\}\left\{i \gamma_0 
{\rm sin}\left(\frac{p_0 a_t + q_0 a_t}{2}\right)
-r {\rm cos}\left(\frac{p_0 a_t + q_0 a_t}{2}\right)\right\},\\
V_{2ii}^{AB}(q,p)&=&\frac{a_s}{2} g^2 \frac{1}{2}
\left\{T^A,T^B\right\} \left\{\nu i \gamma_i 
{\rm sin}\left(\frac{p_i a_s + q_i a_s}{2}\right)
-\frac{r\eta}{\xi} 
{\rm cos}\left(\frac{p_i a_s + q_i a_s}{2}\right)\right\},\\
V_{c0}^{A}(q,p)&=&-\ce g T^A \frac{r(1+\eta)}{4\xi}
\sum_i \sigma_{0i}{\rm sin}\left(p_i a_s - q_i a_s \right)
{\rm cos}\left(\frac{p_0 a_t - q_0 a_t}{2}\right),\\
V_{ci}^{A}(q,p)&=&-\ce g T^A \frac{r(1+\eta)}{4}
\sigma_{i0}{\rm sin}\left(p_0 a_t - q_0 a_t \right)
{\rm cos}\left(\frac{p_i a_s - q_i a_s}{2}\right)\nn\\
&&-\cb g T^A \frac{r \eta}{2\xi}
\sum_{j} \sigma_{ij}{\rm sin}\left(p_j a_s - q_j a_s \right)
{\rm cos}\left(\frac{p_i a_s - q_i a_s}{2}\right),
\een
where $p$ is for the incoming momentum into the vertex and
$q$ for the outgoing momentum. $T^A$ $(A=1,\dots,N_c^2-1)$
is a generator of color SU($N_c$).

For the gauge part we take the standard Wilson action on the
anisotropic lattice:
\ben
S_g&=&\frac{2 N_c}{g^2}\sum_n \left[
\xi \sum_{i}\left(1-\frac{1}{2N_c} {\rm Tr}(U_{0i}(n)
+U_{0i}^\dagger(n))\right)\right. \nn\\
&&\;\;\;\;\left. +\frac{1}{\xi} \sum_{i<j}\left(1-
\frac{1}{2N_c} {\rm Tr}(U_{ij}(n)+U_{ij}^\dagger(n))\right)\right]
\een 
with
\be
U_{\mu\nu}(n)=U_{\mu}(n)
U_{\nu}(n+{\hat \mu})U_{\mu}^\dagger(n+{\hat \nu})
U_{\nu}^\dagger(n).
\ee
The gluon propagator is given by
\be
G_{\mu\nu}^{AB}(k)=a_s^2\frac{\delta_{\mu\nu}\delta_{AB}}
{\xi^2 4 {\rm sin}^2\left(\frac{k_0 a_t}{2}\right)
+4\sum_i {\rm sin}^2\left(\frac{k_i a_s}{2}\right)
+\lambda^2a_s^2}
\ee
in the Feynman gauge with the fictitious mass $\lambda^2$,
which is introduced to work as the infrared cutoff. 

\subsection{Tree-level analysis on the quark propagator}

Expanding the inverse free quark propagator in
eq.~(\ref{eq:qp_tree}) up to $O(a_t)$ we obtain 
\be
S_q^{-1}(p)=i \gamma_0 p_0+\nu i \sum_i \gamma_i p_i + m_0
+\frac{r a_t}{2} p_0^2 + \frac{r\eta a_t}{2} \sum_i p_i^2
+O((p_0 a_t)^2,(p_i a_s)^2) ,
\ee
which yields the following expression for the quark propagator: 
\be
S_q(p)=\frac{1}{1+m_0 r a_t}
\frac{-i \gamma_0 p_0-\nu i \sum_i \gamma_i p_i + m_0
+\frac{r a_t}{2}\left( 
p_0^2 + \eta \sum_i p_i^2
\right)}
{p_0^2+\frac{\nu^2+m_0 r\eta a_t}{1+m_0 r a_t}\sum_i
p_i^2 + \frac{1}{1+m_0 r a_t}m_0^2 } + O(a_{t,s}^2) .
\label{eq:qprop}
\ee
We consider 
the tree-level on-shell improvement for this quark
propagator requiring that 
$S_q(p)$ should reproduce the form 
\be
S_q(p)=
\frac{1}{Z_q}
\frac{ -i \gamma_0 p_0- i \sum_i \gamma_i p_i + m_R}
{p_0^2+\sum_i p_i^2 +m_R^2}
+{\rm (no\; pole\; terms)}+O(a_{t,s}^2)
\label{eq:qprop_r}
\ee
with the appropriate choice for $Z_q$, $Z_m$, 
$\nu$ and $\eta$, where
$Z_q$ and $Z_m$ denote renormalization factors
for the quark wave function and mass defined by
\ben
q_R&=&Z_q^{1/2} q,\\
m_R&=&Z_m m_0,
\een
where $m_R$ is the pole mass.
It should be reminded that $r$ is a free parameter.
We remark that terms without the pole in the quark
propagator of eq.~(\ref{eq:qprop_r}) yield only contact
terms in the configuration space, which do not contribute to
on-shell quantities in Green functions.
In terms of the inverse quark propagator,
the condition eq.~(\ref{eq:qprop_r}) is equivalent to
\ben
S^{-1}_q(p) &=& \left[ Z_q - 2 C m_0 a_t \right]( i \slas{p}{} + m_R)
+ C a_t (p^2+m_R^2) +O( a_t(i\slas{p}{} + m_R)^2) + O(a_t^2)
\een
with $C$ constant.
Therefore ``(no pole terms)'' in eq.~(\ref{eq:qprop_r}) 
are not necessary to be $O(a_t^2)$.
Comparing the expressions of  
eqs.~(\ref{eq:qprop}) and (\ref{eq:qprop_r}), we find at the tree level
\ben
Z_q^{-1/2}&=&1-\frac{r}{2}m_0 a_t,\\
Z_m&=&1-\frac{r}{2}m_0 a_t,\\
\nu&=&1,\\
\eta&=&1.
\label{eq:condition}
\een

Up to now three types of quark actions with different choices for
$r$ and $\eta$ have been proposed: 
(i) $r=1$ and $\eta=1$\cite{symm_at},
(ii) $r=\xi$ and $\eta=1$\cite{symm_as} and
(iii) $r=1$ and $\eta=\xi$\cite{pp}.
Although the action in the case of (iii) has been most extensively studied 
numerically, the choice of the parameter $\eta$ does not
meet the condition of eq.~(\ref{eq:condition}) that is
required from the on-shell improvement at the tree-level.
This primitive failure makes us consider that it is not worthwhile to
work on the case (iii) in this paper.
We focus on only cases (i) and (ii) hereafter.

Let us first derive the relation between the bare quark mass $m_0$ and 
the pole mass $m_p$.
Putting $p_i=0$ and $p_0=i m_p$
into the inverse free quark propagator of eq.~(\ref{eq:qp_tree}),
the on-shell condition yields
\be
m_p a_t = {\rm log}\left| \frac{m_0 a_t +r 
+\sqrt{(m_0 a_t)^2+2 r m_0 a_t +1}}{1+r} \right|. 
\ee
While in the case (i) with $r=1$ we can expand 
$m_p a_t$ in powers of
$m_0$ under the condition $m_0 a_t \ll 1$,
in the case (ii) with $r=\xi$ the condition $\xi m_0 a_t=m_0 a_s\ll 1$
is necessary. To avoid any confusions we assume  
$m_0 a_s\ll 1$ from now on. We remark that this
assumption does not affect any conclusions in this section.

On the anisotropic lattice we have to be careful about 
contributions of space doublers.
Pole masses of space doublers are written as
\be
m_p^d a_t = {\rm log}\left| \frac{m_0^d a_t +r 
+\sqrt{(m_0^d a_t)^2+2 r m_0^d a_t +1}}{1+r} \right|
\ee
with
\be
m_0^d=m_0+\frac{2 r }{\xi^2} \frac{N_d}{a_t},
\ee
where $N_d$ components of spatial momentum 
$p_i$ are equal to $\pi/a_s$ at the edge of the Brillouin
zone. Although doubler pole masses are always heavier than
the physical one irrespective of the value of $m_0$, $r$ and
$\xi$, their differences in the large limit of $\xi$ are
given by
\ben
(m_p^d-m_p)a_s &\rightarrow& 
\frac{2}{\xi} N_d \frac{1}{1+m_0 a_t}+O\left(\frac{1}{\xi^2}\right)
\;\;\;\; {\rm case\; (i)},\\
(m_p^d-m_p)a_s &\rightarrow& 
\sqrt{1+2 m_0 a_s+4 N_d}-\sqrt{1+2 m_0 a_s}+O\left(\frac{1}{\xi}\right)
\;\;\;\; {\rm case\; (ii)}.
\een
For the case (i) we find the gap $(m_p^d-m_p)a_s$ diminishes as
$\xi$ becomes larger. This brings a
practical problem in numerical studies of heavy quarks:
contributions of doublers could contaminate 
signals of hadron states.
On the other hand, we are free from this problem 
in the case (ii).

\subsection{One-loop quark self energy}

\begin{figure}[h]
\centering{
\hskip -0.0cm
\psfig{file=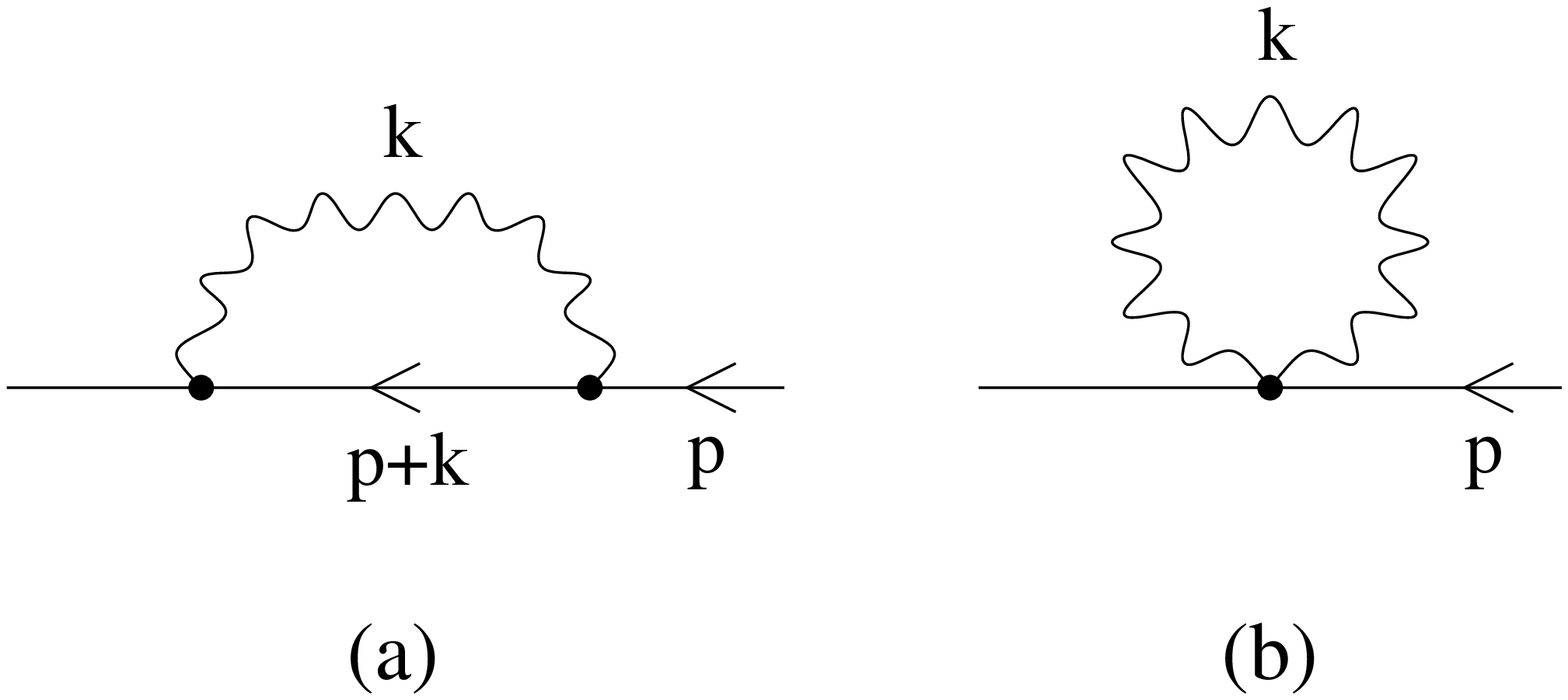,width=140mm,angle=-90}
\vskip +3mm 
}
\caption{One-loop diagrams for the quark self energy.} 
\label{fig:qse}
%\vspace{-8mm}
\end{figure}

The inverse full quark propagator is written as
\be
S_q^{-1}(p)=i \gamma_0 p_0+\nu i \sum_i \gamma_i p_i + m_0
+\frac{a}{2} p_0^2 + \eta \frac{a}{2} \sum_i p_i^2-\Sigma(p,m_0),
\label{eq:qp_inv}
\ee
where we take $a=r a_t$(;$a=a_t$ for (i) while $a=a_s$ for (ii)).
One-loop contributions to the quark self-energy
$\Sigma(p,m_0)$ consist of two types of diagrams depicted 
in Figs.~\ref{fig:qse} (a) and (b), which are expressed by
\ben
\Sigma_a(p,m)&=&\int \frac{d^4 k}{(2\pi)^4}\sum_A\sum_{\mu}
\left\{
V_{1\mu}^A(p,p+k) S_q(p+k) V_{1\mu}^A(p+k,p)\right. \nn\\
&&\;\;\;\;\left.+V_{c\mu}^A(p,p+k) S_q(p+k) V_{1\mu}^A(p+k,p)\right. \nn\\ 
&&\;\;\;\;\left.+V_{1\mu}^A(p,p+k) S_q(p+k) V_{c\mu}^A(p+k,p)\right. \nn\\
&&\;\;\;\;\left.+V_{c\mu}^A(p,p+k) S_q(p+k) V_{c\mu}^A(p+k,p) 
\right\}G_{\mu\mu}^{AA}(k)
\een
and
\be
\Sigma_b(p)=\int \frac{d^4 k}{(2\pi)^4}\sum_A\sum_{\mu}
V_{2\mu\mu}^{AA}(p,p) G_{\mu\mu}^{AA}(k)
\ee
with 
\ben
&&-\frac{\pi}{a_t}\le k_0 \le \frac{\pi}{a_t},\\
&&-\frac{\pi}{a_s}\le k_i \le \frac{\pi}{a_s}.
\een  
Expanding $\Sigma(p,m_0)=\Sigma_a(p,m_0)+\Sigma_b(p)$ in terms
of $p$ and $m_0$, we obtain  
\ben
\Sigma(p,m_0)&=&\frac{g^2}{16\pi^2} C_F\left[ \frac{\Sigma_0}{a}
+i \gamma_0 p_0(-L+\Sigma_1^t)+i \sum_i \gamma_i p_i(-L+\Sigma_1^s)
+m_0(-4L+\Sigma_2)\right. \nn\\
&&\;\;\;\;\left. +ap_0^2((1-3c_{\rm SW})L/2+\sigma_1^t)
+a\sum_i p_i^2((1-3c_{\rm SW})L/2+\sigma_1^s)\right.\nn\\
&&\;\;\;\;\left. +am_0 i \gamma_0 p_0((5+3c_{\rm SW})L/2+\sigma_2^t)
+am_0 i \sum_i \gamma_i p_i((5+3c_{\rm SW})L/2+\sigma_2^s)\right.\nn\\
&&\;\;\;\;\left. +am_0^2((5-3c_{\rm SW})L+\sigma_3)\right]+O(a_{t,s}^2)
\label{eq:qse}
\een
with $L=-{\rm log}(\lambda^2 a_s^2)$ 
the contribution of the infrared divergence.
Here we take $c_B = c_E \equiv c_{\rm SW}$ for the clover coefficient.
$\Sigma_0$, $\Sigma_1^{t,s}$, $\Sigma_2$ and 
$\sigma_1^{t,s}$, $\sigma_2^{t,s}$, $\sigma_3$ are independent of
$a$ but functions of $\xi$, $\eta$  and $r$ parameters.
We evaluate these quantities numerically using the Monte
Carlo integration routine BASES\cite{bases}.

 From eqs.~(\ref{eq:qp_inv}) and (\ref{eq:qse}) we obtain
\ben
Z_q^{-1/2}&=&\left\{1+\frac{g^2}{16\pi^2}
\frac{C_F}{2}(-L+\Delta_q^{(0)})\right\}
\left\{1+a_t m\left(-\frac{r}{2}+\frac{g^2}{16\pi^2}C_F\Delta_q^{(1)} 
\right)\right\},\\
Z_m&=&\left\{1+\frac{g^2}{16\pi^2} C_F(3L+\Delta_m^{(0)}\right\}
\left\{1+a_t m\left(-\frac{r}{2}+\frac{g^2}{16\pi^2} C_F\Delta_m^{(1)}
\right)\right\},\\
\nu&=&1-\frac{g^2}{16\pi^2} C_F\Delta_\nu^{(0)}
-\frac{g^2}{16\pi^2} C_F a_t m \Delta_\nu^{(1)},\\
\eta&=&1-2 \frac{g^2}{16\pi^2} C_F \Delta_\eta^{(0)},
\een 
where
\ben
m&=&m_0-\frac{g^2}{16\pi^2}C_F\frac{\Sigma_0}{a},\\
\Delta_q^{(0)}&=&\Sigma_1^t,\\
\Delta_q^{(1)}&=&r\left(\sigma_1^t+
\frac{\sigma_2^t}{2}-\Sigma_1^t+\frac{\Sigma_2}{2}
+\frac{3(1-c_{\rm SW})}{4}L
\right),\\ 
\Delta_m^{(0)}&=&\Sigma_1^t-\Sigma_2,\\
\Delta_m^{(1)}&=&r\left(\sigma_1^t+
\sigma_2^t-\sigma_3-\Sigma_1^t+\frac{\Sigma_2}{2}
+3(c_{\rm SW}-1)L
\right),\\
\Delta_\nu^{(0)}&=&\Sigma_1^t-\Sigma_1^s,\\
\Delta_\nu^{(1)}&=&r \left(\sigma_2^t-\sigma_2^s\right),\\
\Delta_\eta^{(0)}&=&\sigma_1^t-\sigma_1^s.
\een
We find for the anisotropic case that 
the $g^2a {\log} a$ terms disappear in $Z_q$, $Z_m$ for $c_{\rm SW}$ =1 
as well as $\nu$ and $\eta$ for an arbitrary values of $c_{\rm SW}$,
as observed in $Z_q$ and $Z_m$ for the isotropic case\cite{sw_pt}.
Thus $c_{\rm SW}=1$ gives the tree level estimate for $c_B$ and $c_E$,
and we take this value for the latter numerical calculation in this section.
With this choice for $Z_q$, $Z_m$, $\nu$ and $\eta$ the
quark propagator is given by
\ben
S_q(p)&=&\frac{Z_q^{-1}}{p_0^2+\sum_i p_i^2 +m_R^2}
\left[-i \gamma_0 p_0 -i \sum_i \gamma_i p_i + m_R\right.\nn\\
&&\;\;\;\;\left. +a(p_0^2+\sum_i p_i^2 +m_R^2)\left\{\frac{1}{2}
-\frac{g^2}{16\pi^2}C_F
\left(-\frac{L}{2}+\sigma_1^t-\frac{\Sigma_1^t}{2}\right)\right\}\right]\\
&=&Z_q^{-1} S_q^R(p)+Z_q^{-1}a\left\{\frac{1}{2}
-\frac{g^2}{16\pi^2}C_F
\left(-\frac{L}{2}+\sigma_1^t-\frac{\Sigma_1^t}{2}\right)\right\},
\een
where $S_q^R(p)$ is the renormalized quark propagator.
We again remark that the term without the pole in the quark
propagator does not contribute to
on-shell quantities in Green functions.

\begin{figure}[h]
\centering{
\hskip -0.0cm
\psfig{file=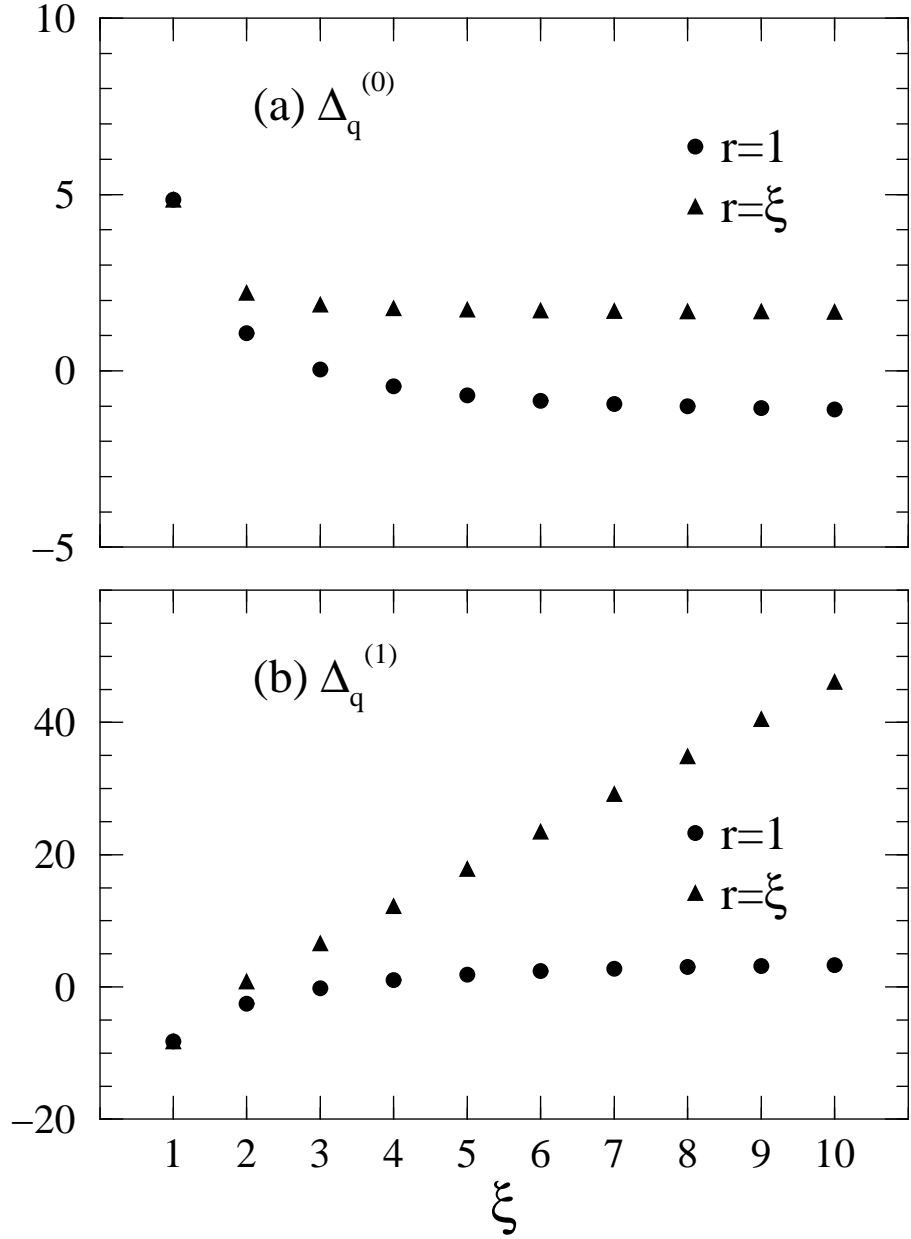,width=120mm,angle=0}
\vskip +3mm 
}
\caption{$\xi$ dependences of (a) $\Delta_q^{(0)}$ 
and (b) $\Delta_q^{(1)}$ in
the renormalization constant of the quark wave function.
$\ce$ and $\cb$ are chosen to be 1. Errors are within symbols.} 
\label{fig:zwf}
%\vspace{-8mm}
\end{figure}

\begin{figure}[h]
\centering{
\hskip -0.0cm
\psfig{file=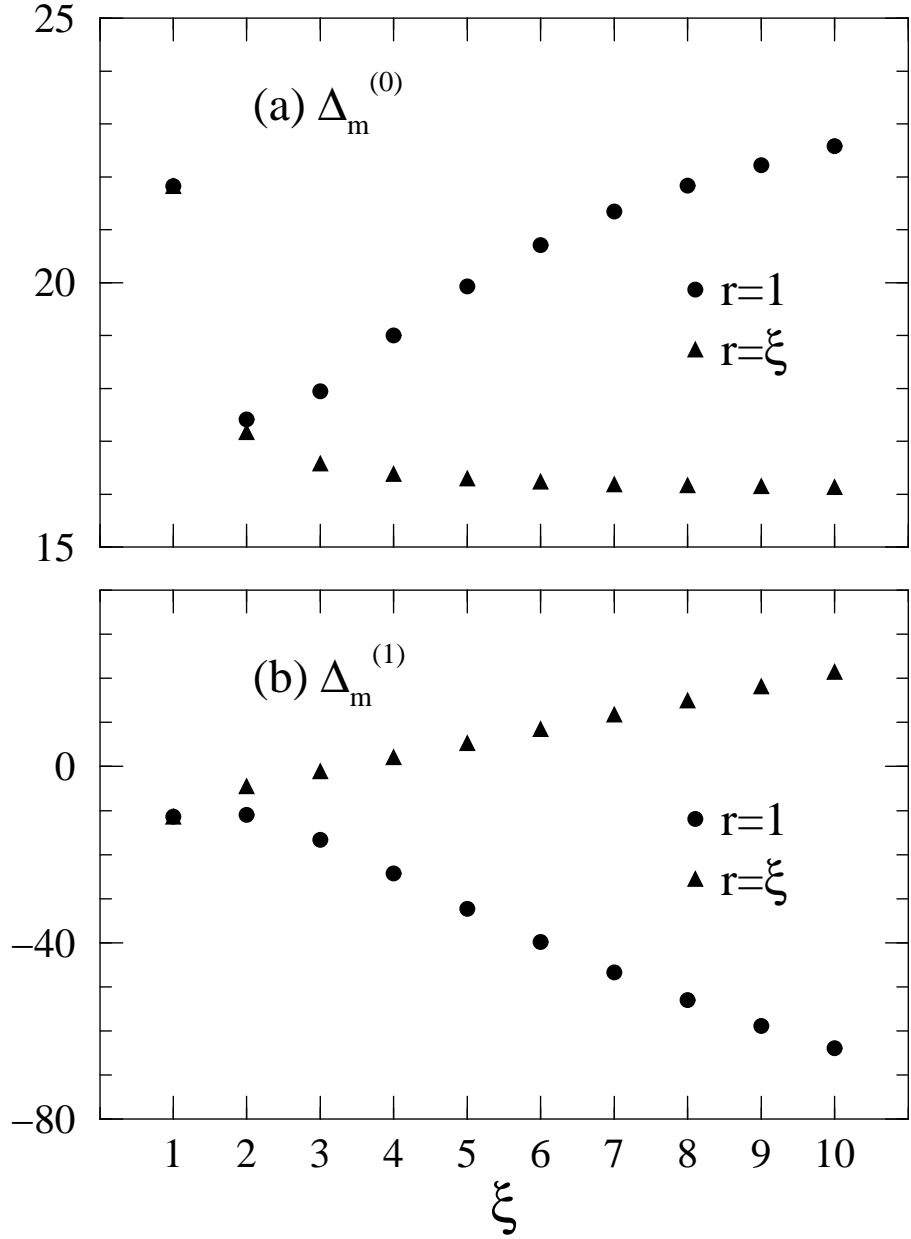,width=120mm,angle=0}
\vskip +3mm  
}
\caption{$\xi$ dependences of (a) $\Delta_m^{(0)}$ 
and (b) $\Delta_m^{(1)}$ in
the renormalization constant of the quark mass.
$\ce$ and $\cb$ are chosen to be 1. Errors are within symbols.}
\label{fig:zm}
%\vspace{-8mm}
\end{figure}

\begin{figure}[h]
\centering{
\hskip -0.0cm
\psfig{file=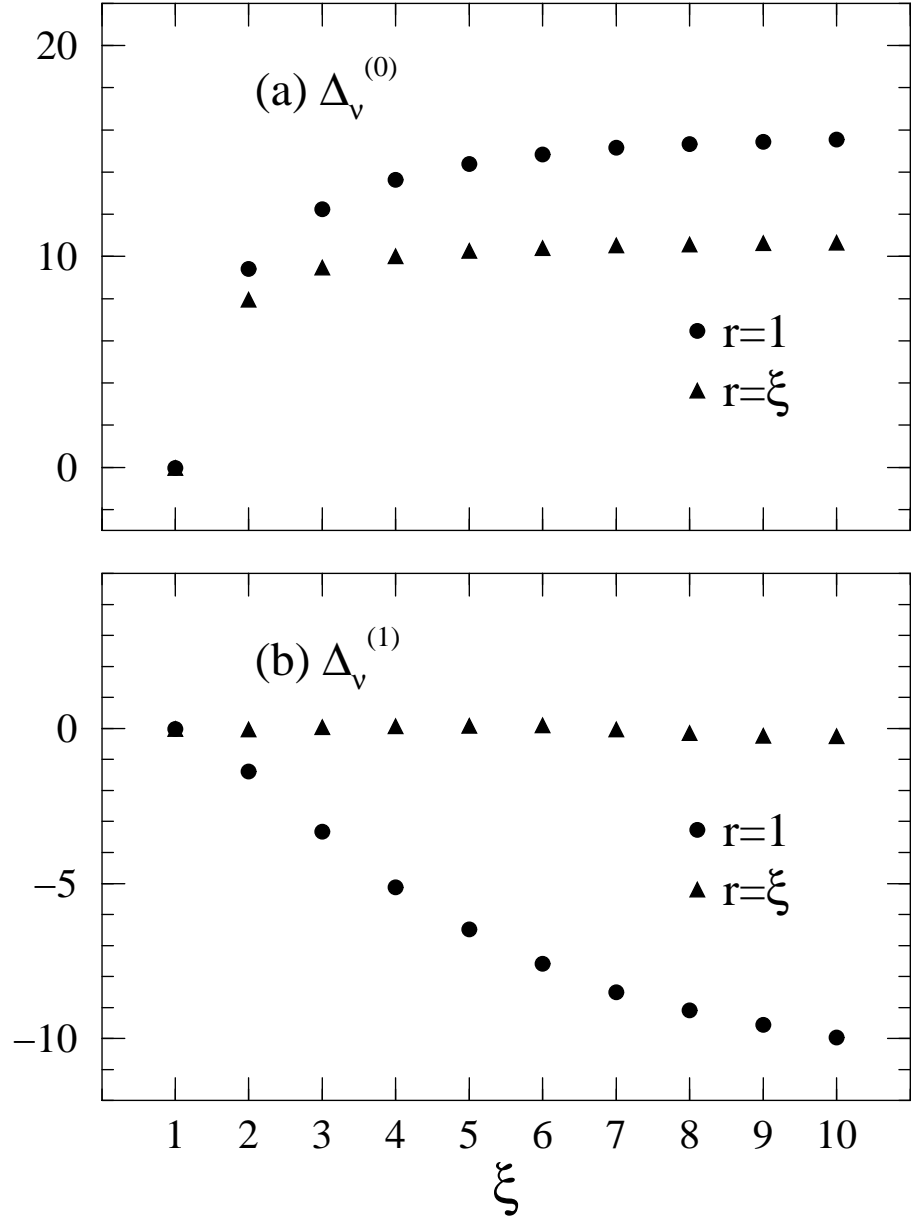,width=120mm,angle=0}
\vskip +3mm 
}
\caption{$\xi$ dependences of (a) $\Delta_\nu^{(0)}$ 
and (b) $\Delta_\nu^{(1)}$ in
the $\nu$ parameter.
$\ce$ and $\cb$ are chosen to be 1. Errors are within symbols.}
\label{fig:nu}
%\vspace{-8mm}
\end{figure}

\begin{figure}[h]
\centering{
\hskip -0.0cm
\psfig{file=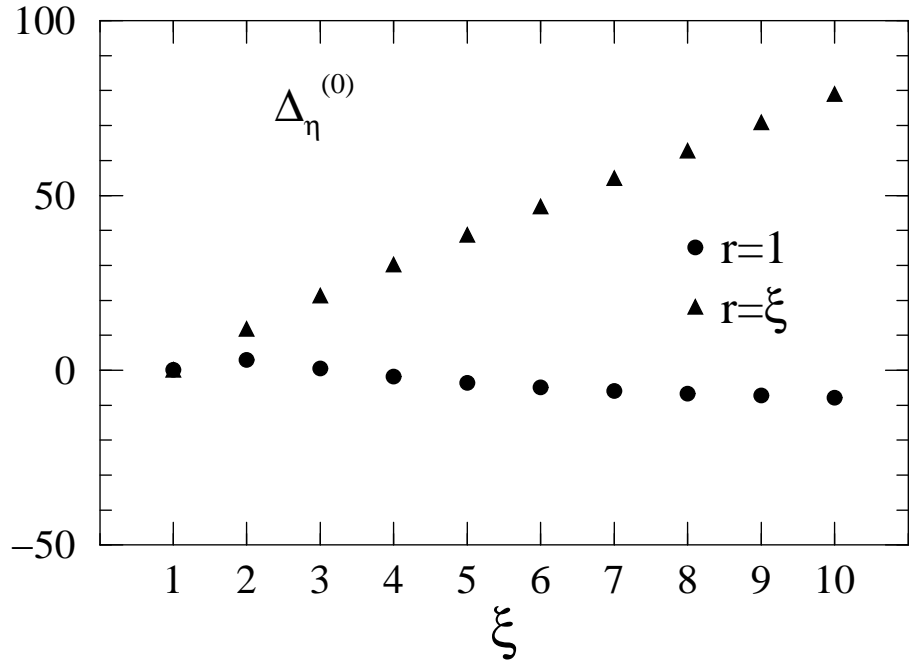,width=120mm,angle=0}
\vskip +3mm 
}
\caption{$\xi$ dependences of $\Delta_\eta^{(0)}$ 
in the $\eta$ parameter.
$\ce$ and $\cb$ are chosen to be 1. Errors are within symbols.}
\label{fig:eta}
%\vspace{-8mm}
\end{figure}

We show $\xi$ dependences of one-loop coefficients
for $Z_q$, $Z_m$, $\nu$ and $\eta$ in Figs.~\ref{fig:zwf}, 
\ref{fig:zm}, \ref{fig:nu} and \ref{fig:eta}, respectively.
Here we consider both $r=1$ and $r=\xi$ cases with $\eta=1$, which satisfy
the tree-level on-shell condition in eq.~(\ref{eq:condition}).
Although our main concern is whether $m a_s$ corrections
could revive at the one-loop level 
for the $r=1$ case, we also present the results of the 
$r=\xi$ case for comparison.
Results for $\Delta_q^{(0)}$, $\Delta_q^{(1)}$, $\Delta_m^{(0)}$ and 
$\Delta_m^{(1)}$  in the isotropic case are already given in
Refs.~\cite{pt_g2,pt_g2a}.
They show an agreement with our results with the choice of $\xi=1$.
For the $r=1$ case, Figs.~\ref{fig:zm}(b),
\ref{fig:nu}(b) and \ref{fig:eta} show
approximate linear dependences on $\xi$ for
$\Delta_m^{(1)}$, $\Delta_\nu^{(1)}$ and
$\Delta_\eta^{(0)}$, which tells us that 
$O(g^2 a)$ contributions to $Z_m$, $\nu$ and $\eta$ 
are effectively of order $g^2 m
a_s=g^2 m a_t\xi$.   
We observe similar linear dependences for the $r=\xi$ case
in Figs.~\ref{fig:zwf}(b),
\ref{fig:zm}(b) and \ref{fig:eta}.
 From these observations we conclude that $m a_s$
corrections are allowed to revive at the one-loop level.

This is a reasonable conclusion in view of the on-shell
improvement. As far as we know there is no symmetry on the
anisotropic lattice to prohibit the higher dimensional operators
with the form of $(m a_s)^n {\cal O}_{4+k}$ ($n,k\ge 1$), 
where ${\cal O}_{4+k}$ denotes $4+k$ dimensional operators.
Unless such symmetry is uncovered, the theoretical advantage of
the anisotropic lattice over the isotropic one would 
be never confirmed.

\section{Isotropic lattice}
\label{sec:iso}

In this section we propose a new relativistic approach to
control $m_Q a$ corrections for the heavy quarks on the
lattice. The idea is based on the on-shell improvement
programme applied to the heavy quarks 
on the isotropic lattice. 
This method allows us to obtain the physical quantities 
in the continuum limit   
without requiring harsh condition $m_Q a \ll 1$ that is not 
achievable in near future.

Let us consider the general cutoff effects for the heavy quarks on
the lattice.
Here we assume that the heavy quark mass $m_Q$ is much
heavier than $\Lambda_{\rm QCD}$, while the light quark mass $m_q$
is lighter than $\Lambda_{\rm QCD}$.
Under this condition we assume that the leading cutoff
effects are
\be
f_0(m_Q a)>f_1(m_Q a)a\lqcd>f_2(m_Q a)(a\lqcd)^2>\cdots,
\ee
where $f_i(m_Q a)$ ($i\geq 0$) are smooth and continuous 
all over the range of 
$m_Q a$ and have Taylor expansions at $m_Q a=0$ with
sufficiently large convergence radii beyond $m_Q a=1$.
To control the scaling
violation effects we want to remove the cutoff effects up to
$f_1(m_Q a)a\lqcd$ by adding the counter terms
to the lattice quark action with the on-shell improvement.
If $m_Q a$ is small enough,
the remaining $f_2(m_Q a)(a\lqcd)^2$ contributions can be removed by
extrapolating the numerical data at several lattice spacings 
to the continuum limit. Otherwise, in case of
sufficiently small lattice spacing, 
the $O((a\Lambda_{\rm QCD})^2)$ errors can be neglected.
Our aim in this section is to search for the relevant counter terms
required in the on-shell improvement and propose a
nonperturbative method to determine their coefficients. 
We also show that the behavior of $f_i(m_Q a)$ ($i\geq 1$) in
the large $m_Q a$ region can be discussed by investigating the 
static limit.

\subsection{On-shell improvement on the isotropic lattice}
\label{subsec:onshell_iso}

We first list the allowed operators under the
requirement of the gauge, axis permutation and other various
discrete symmetries on the lattice, where
the chiral symmetry is not imposed.
According to the work of Ref.~\cite{sw}, all the operators
with dimension up to six are given by
\ben
{\rm dim.3:}&& {\cal O}_3(x)={\bar q}(x)q(x), \\ 
{\rm dim.4:}&& {\cal O}_{4}(x)={\bar q}(x)\slal{D}{} q(x),\\
{\rm dim.5:}&& {\cal O}_{5a}(x)={\bar q}(x)D_\mu^2 q(x),\\
            && {\cal O}_{5b}(x)=i{\bar q}(x)
                                \sigma_{\mu\nu}F_{\mu\nu} q(x),\\
{\rm dim.6:}&& {\cal O}_{6a}(x)={\bar q}(x)\gamma_\mu D_\mu^3 q(x),\\
            && {\cal O}_{6b}(x)={\bar q}(x)D_\mu^2 \slal{D}{} 
                                q(x),\\
            && {\cal O}_{6c}(x)={\bar q}(x)\slal{D}{} 
                                D_\mu^2 q(x),\\
            && {\cal O}_{6d}(x)=i{\bar q}(x)\gamma_\mu[D_\nu,
                                F_{\mu\nu}] q(x),\\
            && {\cal O}_{6e}(x)={\bar q}(x){\slal{D}{}}^{\,3} q(x),\\
            && {\cal O}_{6f}(x)={\bar q}(x)\Gamma q(x)
                                {\bar q}(x)\Gamma q(x),
\een  
where $\Gamma=1,\gamma_5,\gamma_\mu,\gamma_\mu\gamma_5,\sigma_{\mu\nu}$.
The higher dimensional operators are
related to the lower dimensional operators with additional
$(m_Q a)^n$ corrections with and without classical field
equations, or otherwise they have the contributions 
of order $(a\Lambda_{\rm QCD})^2$ or less as the cutoff effects.

These operators lead to a following generic form of the quark
action on the isotropic lattice:
\be
S_q^{\rm imp}=\sum_x\left[ 
c_3{\cal O}_3(x)+c_{4}{\cal O}_{4}(x) 
+\sum_{i=a,b} c_{5i}{\cal O}_{5i}(x) 
+\sum_{i=a,\dots,f} c_{6i}{\cal O}_{6i}(x)  
\right],
\label{eq:qimp_iso}
\ee
where $c_3,\dots,c_{6f}$ are functions of the bare gauge
coupling $g$ and the power corrections of $ma$.
We first remark that the $m a$ corrections to the quark mass term
and the kinetic term can be absorbed 
in the renormalizations of the quark mass $Z_m$ and the
wave function $Z_q$.
For the sake of convenience we choose $c_3=m_0$ and $c_4=1$.

In the next step we reduce the number of basis
operators with the aid of the classical field equations.
It is easily found that ${\cal O}_{5a}$, ${\cal O}_{6b}$, 
${\cal O}_{6c}$ and  ${\cal O}_{6e}$ can be 
related to the quark mass term
or the kinetic term. In the on-shell improvement these
operators are redundant and can be eliminated 
from the action of eq.~(\ref{eq:qimp_iso}).
The operator ${\cal O}_{5a}$, however, 
is used to avoid the species doubling and the value of its
coefficient $c_{5a}$ is given by hand.  

The remaining operators are
${\cal O}_{5b}$, ${\cal O}_{6a}$, ${\cal O}_{6d}$ and 
${\cal O}_{6f}$,
whose contributions as the cutoff effects  
are estimated in Table~\ref{tab:action}.
We use the classical field equation for 
${\bar q}(x)\gamma_0 D_0^3 q(x)$ in ${\cal O}_{6a}$.
Here it should be noted that $\lqcd$ means 
the order $\Lambda_{\rm QCD}$ or less throughout this paper.
In some cases the actual contribution may become smaller.
For example,
the contributions of the bilinear terms 
whose Dirac matrices consist of off-diagonal components
are suppressed by the extra $\Lambda_{\rm QCD}/m_Q$
for the heavy quarks compared to the light quarks.
The operator ${\cal O}_{5b}$ is the so-called clover term,
for which
the nonperturbative method to determine the coefficient 
$c_{5b}$ in the massless limit is already
established\cite{impr}. 
However, the contributions of $(ma)^n{\cal O}_{5b}$ 
($n\geq 1$) cannot be neglected in the 
present condition that allows $m_Q a\sim O(1)$. 
For $O(a\Lambda_{\rm QCD})$ improvement
the coefficient $c_{5b}$ has to be adjusted in the mass dependent way.
The differences in magnitude between the time and space components in
${\cal O}_{6a}$ 
originate from the violation of rotational
symmetry on the lattice with finite lattice spacing.
While the contributions of the space components 
are found to be negligible, 
those of the time components should be removed. 
We also find the contributions of the four-quark operators
in ${\cal O}_{6f}$ are negligible.

The generalization of 
the above argument to any operators with higher
dimensions makes the discussion more transparent.
Let us consider  an arbitrary operator with
$4+k$ dimension,
$a^k {\cal O}_{4+k}$, where we write the lattice spacing $a$ explicitly.
The operator ${\cal O}_{4+k}$ contains $l$ pairs of $\bar q$ and $q$ and
$n$ covariant derivatives $D_\mu$ with $4+ k = 3 \times l + n$.
Using the classical field equation, 
some (but not all) of covariant derivatives
can be replaced by the quark mass $m$. For $l\ge 2$ 
the largest possible power of the scaling violation is 
$ (m a )^n (a\Lambda_{\rm QCD})^{3l-4}$. 
Therefore the operators which contain four or more quarks
are irrelevant for the $O(a\Lambda_{\rm QCD})$ improvement.
All the relevant contributions come from the quark bilinear operators. 
With the aid of the classical field equations, they can be
reduced to
\ben
&& (m a)^n a^{-1} {\bar q}(x)q(x)  \\
&& (m a)^{n-1} {\bar q}(x)\gamma_0 D_0 q(x),\;\; 
(m a)^{n-1} \sum_i {\bar q}(x)\gamma_i D_i q(x)\\
&& (m a)^{n-2}a {\bar q}(x) D_0^2 q(x),\;\; 
(m a)^{n-2}a \sum_i {\bar q}(x) D_i^2 q(x)\\
&& (m a)^{n-2} a i\sum_i {\bar q}(x)\sigma_{0i}F_{0i} q(x),\;\;
(m a)^{n-2} a i\sum_{ij} {\bar q}(x)\sigma_{ij}F_{ij} q(x),
\een
for $n \geq 0$. The time and space components of 
${\cal O}_{4}$ and ${\cal O}_{5a,5b}$ should be treated separately
in case of finite $ma$, where
the space-time asymmetry reflects the contributions of the higher
dimensional operators that break the rotational symmetry.
Now we know that the seven operators are
needed for the $O(a\lqcd)$ improvement.
Since three coefficients among these seven operators
can be absorbed in $Z_m$, $Z_q$ and the Wilson parameter $r_t$ 
for the time derivative as already explained,
the remaining four coefficients have to be actually tuned.

In conclusion, at all order of $m a$,
the generic quark action is
written as
\ben
S_q^{\rm imp}&=&\sum_x\left[ m_0{\bar q}(x)q(x)
+{\bar q}(x)\gamma_0 D_0q(x)
+\nu \sum_i {\bar q}(x)\gamma_i D_i q(x)
-\frac{r_t a}{2} {\bar q}(x)D_0^2 q(x)\right.\nn\\
&&\;\;\;\;\;\;\;\;\left.-\frac{r_s a}{2} \sum_i {\bar q}(x)D_i^2 q(x)
-\frac{ig a}{2}c_E \sum_i {\bar q}(x)\sigma_{0i}F_{0i} q(x)
-\frac{ig a}{4}c_B \sum_{i,j} {\bar q}(x)\sigma_{ij}F_{ij} q(x)
\right],
\label{eq:qaction_iso}
\een
where we are allowed to choose $r_t=1$ and the four parameters
$\nu$, $r_s$, $c_E$ and $c_B$ are to be adjusted.
In general these parameters have the form that
$X=\sum_n X_n(g^2) (m a)^n$ 
with $X=\nu$, $r_s$, $c_E$ and $c_B$, and
$X_0$ should agree with the one in the massless $O(a)$ improved
theory: $\nu_0=1$, $(r_s)_0=r_t=1$, $(c_E)_0=(c_B)_0 = c_{\rm SW}$\cite{impr}.
Note that $\nu = 1 + O( (ma)^2)$ and $r_s = r_t + O( ma)$ since
the space-time asymmetry arises from Lorentz non-covariant terms such as
${\cal O}_{6a}$ via the on-shell reduction,
accompanied by extra $(ma)^2$ factors.

{}From the above consideration, 
the leading scaling violation in the massive theory,
except for 
$\sum_{n=1}^\infty C_n^{q,m}(g^2, \log a) (m_Q a)^n$ in $Z_q$ and $Z_m$,
is $\sum_{n=0}^\infty C_n^{\rm W}(g^2, \log a) (m_Q a)^n a\Lambda_{\rm QCD} $
for the Wilson quark action, or 
$\sum_{n=1}^\infty C_n^{\rm SW}(g^2, \log a) (m_Q a)^n 
a\Lambda_{\rm QCD} $ for the (massless) $O(a)$ improved SW quark action.
Here we should notice that the contribution of 
$(\nu-1) \sum_i {\bar q}(x)\gamma_i D_i q(x)$ 
is of order $m_Q^2 a^2\Lambda_{\rm QCD}/m_Q\sim m_Q a^2\Lambda_{\rm QCD}\sim
a\Lambda_{\rm QCD}$ in the heavy quark region.
This implies that once we fix the pole mass from some spectral
quantity, the cutoff effects in other spectral quantities are
at most of order $a\Lambda_{\rm QCD}$, not $(m_Q a)^n$, 
for the Wilson and the
(massless) $O(a)$ improved SW quark actions.
It should be noted that the quark wave function does not affect on the
spectral quantities.
If $\nu$, $r_s$, $c_E$ and $c_B$ are
properly adjusted in the mass dependent way,
the remaining scaling violations are reduced to 
$\sum_{n=0}^\infty C_n^{\rm ours}(g^2, \log a) (m_Q a)^n 
(a \Lambda_{\rm QCD} )^2$. 

This relativistic argument about the on-shell improvement
on the massive quarks helps us understand some numerical
results for heavy quark physics previously obtained by using
the Wilson and SW quark actions under the condition $m_Q
a\sim O(1)$.
We find a good example in Fig.~2 of Ref.~\cite{jlqcd96},
which compares
the difference between twice of the heavy-light meson mass and 
the heavy-heavy one, $2m_{HL}-m_{HH}$, obtained by using the pole mass
or the kinetic mass defined by $\partial^2
E_{HH}/\partial p_i^2$.
We observe that $2m_{HL}-m_{HH}$ with the pole mass is
consistent with the experimental values within the
ambiguities of order $a\Lambda_{\rm QCD}$ as expected, while 
$2m_{HL}-m_{HH}$ with the kinetic mass gets deviated from the
experimental values further and further as $m_Q a$ becomes larger.
Recall that the authors of Ref.~\cite{fermilab} suggest from a
nonrelativistic point of view that
the kinetic mass should be used for analyses on the
heavy quark quantities.
Our relativistic argument, however, tells us
that the use of the pole mass makes the remaining cutoff 
effects $O(a\Lambda_{\rm QCD})$.
Since the difference between the kinetic mass and the pole
mass is of order $(m_Q a)^2$ at the tree-level, 
the use of the kinetic mass eventually yields unwanted
additional $(m_Q a)^2$ errors. This is the reason why the results 
of $2m_{HL}-m_{HH}$ with the kinetic mass show considerable
deviation from the
experimental values.

At the end of this subsection we have to remark one point. 
The $m a$ corrections in $Z_m$ and $Z_q$, though they are irrelevant
for spectral quantities, become important,
together with other renormalization factors
in case of calculating the quark
masses and the matrix elements of various composite operators.
In Sec.~\ref{subsec:npr} we will show how to
calculate these renormalization factors nonperturbatively 
including the $(m_Q a)^n$ corrections.

\subsection{Improvement of the axial current}
\label{subsec:axial}

The cutoff effects in the correlation functions of local
composite fields are originated from not only the action but
also the composite fields themselves. 
In this subsection we demonstrate the on-shell improvement
on the axial current, which is relevant for calculation
of the heavy-light pseudoscalar meson 
decay constants like $f_B$ and $f_D$.

The axial current in the $N_f$ flavor space is given by
\be
{\cal A}_\mu^a (x)={\bar q}(x)\gamma_\mu\gamma_5\frac{\lambda^a}{2} q(x),
\ee 
where $\lambda^a/2$ ($a=1,\dots,N_f^2-1$) 
are generators of SU($N_f$). 
The improvement of this operator is performed in the same
way as the improvement of the quark action.
After some consideration we find that it is sufficient to
consider only the dimension four operators 
for the $O(a\lqcd)$ improvement:
the higher dimensional operators can be reduced to ${\cal
A}_\mu^a$ or the
dimension four operators multiplied by $(ma)^n$
using the classical field equations, or otherwise
their contributions are of order $(a\lqcd)^2$. 
The requirement of various
symmetries on the lattice allows the following 
dimension four operators:
\ben
{\rm dim.4:}&& ({\cal A}_{4a})_\mu^a(x)={\bar q}(x)\gamma_5 D_\mu 
                              \frac{\lambda^a}{2} q(x)+
                              {\bar q}(x){\overleftarrow D}_\mu \gamma_5  
                              \frac{\lambda^a}{2} q(x),\\
            && ({\cal A}_{4b})_\mu^a(x)={\bar q}(x)
                                \gamma_5 \sigma_{\mu\nu}D_\nu 
                              \frac{\lambda^a}{2} q(x)
                               -{\bar q}(x){\overleftarrow D}_\nu 
                               \sigma_{\mu\nu}\gamma_5 
                              \frac{\lambda^a}{2} q(x),
\een
where we do not take the sum on the index $\mu$.
We find that $({\cal A}_{4b})_\mu^a$ is related to
$({\cal A}_{4a})_\mu^a$ and ${\cal A}_\mu^a$
with the aid of the classical field equations,
which means the operator $({\cal A}_{4b})_\mu^a$
is redundant for the $O(a\lqcd)$ improvement. 

After all the improved axial current is written as
\be
({\cal A}_\mu^a)^{\rm imp}(x)=
{\cal A}_\mu^a(x)+d_\mu(g^2,ma)({\cal A}_{4a})_\mu^a(x).
\ee
The parameter $d_\mu$ has been already calculated in the massless case. 
Perturbative estimate gives
$d_\mu(g^2,ma=0)=-0.00756g^2$\cite{dmu}, which is fairly
small in magnitude.
 From this fact we think that
it would be sufficient to evaluate $d_\mu$ perturbatively
in the massive case. 
The lattice operator $({\cal A}_\mu^a)^{\rm imp}$ is related to
the continuum operator with
the renormalization factor $Z_A$:
\be
({\cal A}_\mu^a)^{\rm con}(x)
=Z_A(g^2,{\rm log}a,ma)({\cal A}_\mu^a)^{\rm imp}(x).
\ee
The power corrections of $ma$ in $Z_A$ need to be under
control to obtain
the heavy-light pseudoscalar meson decay constants defined
in the continuum regularization scheme.
In Sec.~\ref{subsec:npr} we will show how to remove 
the $ma$ corrections in $Z_A$ with a nonperturbative
method.

\subsection{Benefit of chiral symmetry}
\label{subsec:chiral}

Although the above discussions are free from the chiral symmetry,
it is also interesting to look into what can be changed by
the presence of the chiral symmetry.
For the convenience we treat the quark mass matrix in the
$N_f$ flavor space
\be
M=\left( \begin{array}{cccc}
m_1 & 0 & \cdots & 0     \\
0 & m_2 & \cdots & 0     \\
\vdots & \vdots & \ddots & \vdots  \\
0 & 0 & \cdots & m_{N_f} 
\end{array}
\right)
\ee
as a spurious field $\mm$  
which transforms like
\ben
\mm \rightarrow V_R \mm V_L^\dagger,\\
\mm^\dagger \rightarrow V_L \mm^\dagger V_R^\dagger,
\een
under SU$(N_f)_L\times$SU$(N_f)_R$, where $V_L$ and $V_R$ are
elements of the fundamental representation of
SU$(N_f)_L$ and SU$(N_f)_R$ respectively.
In terms of quark fields $q$ and ${\bar q}$, 
SU$(N_f)_L$ and SU$(N_f)_R$ act on the left and right handed
components,
\ben
q_L=\frac{1-\gamma_5}{2}q,\\
{\bar q}_L={\bar q}\frac{1+\gamma_5}{2},\\
q_R=\frac{1+\gamma_5}{2}q,\\
{\bar q}_R={\bar q}\frac{1-\gamma_5}{2},
\een
whose transformation properties are given by
\ben
&&q_L \rightarrow V_L q_L,\\
&&{\bar q}_L \rightarrow {\bar q}_L V_L^\dagger,\\
&&q_R \rightarrow V_R q_R,\\
&&{\bar q}_R \rightarrow {\bar q}_R V_R^\dagger.
\een
We work all the calculations assuming this symmetry and 
at the end of calculations we can choose $\mm=\mm^\dagger=M$.

With this artifice let us consider which operators among
${\cal O}_3,\dots,{\cal O}_{6f}$ in the quark action are allowed 
under SU$(N_f)_L\times$SU$(N_f)_R$ symmetry.
We easily find that the dimension three and five 
operators are not allowed.
As for the dimension six operators, some four-fermi
operators are excluded.
We also observe that the power corrections of $m_Q a$ emerge
as the form 
\be
M^{2n}\cdot ({\rm SU}(N_f)_L\times{\rm SU}(N_f)_R 
\;\;{\rm invariant}\;\;{\rm operators}) 
\label{eq:ma_power}
\ee
with $n\geq 0$, which means
\ben
&&{\bar q}(x)M^{2n+1}q(x),\\
&&{\bar q}(x)M^{2n}\gamma_\mu D_\mu q(x),\\
&&i{\bar q}(x)M^{2n+1}\sigma_{\mu\nu}F_{\mu\nu} q(x)
\een
are allowed, while
\ben
&&{\bar q}(x)M^{2n}q(x),\\
&&{\bar q}(x)M^{2n+1}\gamma_\mu D_\mu q(x),\\
&&i{\bar q}(x)M^{2n}\sigma_{\mu\nu}F_{\mu\nu} q(x)
\een
are forbidden.
This may be advantageous in controlling the cutoff
effects as $m_Q a$ becomes smaller away from one.
Even for the chiral non-invariant quark action such as
the SW quark action, however,
the leading cutoff effects except for $Z_q$ and $Z_m$
are $ (m_Q a)^n a\lqcd $ with
$n\not=0$, which are of the same order as those in 
the chirally symmetric actions,
once the coefficient of ${\cal O}_{5b}$ in
the quark action is nonperturbatively tuned in the
massless limit.

As for the improvement of the axial current,
the similar argument can be applied. The dimension four operators 
$({\cal A}_{4a})_\mu$ and $({\cal A}_{4b})_\mu$ 
are not allowed by the chiral
symmetry and the power corrections of $m_Q a$ are restricted
to the form of eq.~(\ref{eq:ma_power}).
As an example,
\be
{\bar q}(x)M^{2n+1}\gamma_5 D_\mu \frac{\lambda^a}{2} q(x)+
{\bar q}(x)M^{2n+1}{\overleftarrow D}_\mu \gamma_5  
\frac{\lambda^a}{2} q(x),\\
\ee
with $n\geq 0$ are allowed.

\subsection{$m_Q a$ corrections at tree-level}
\label{subsec:tree}

In order to control the $m_Q a$ corrections 
it should be essential to nonperturbatively 
determine the renormalization factors $Z_m$
and $Z_q$ and
the four parameters $\nu$,
$r_s$, $c_E$ and $c_B$. 
However, we think it is instructive to first 
investigate the $m_Q a$ corrections at the tree-level.

\begin{figure}[h]
\centering{
\hskip -0.0cm
\psfig{file=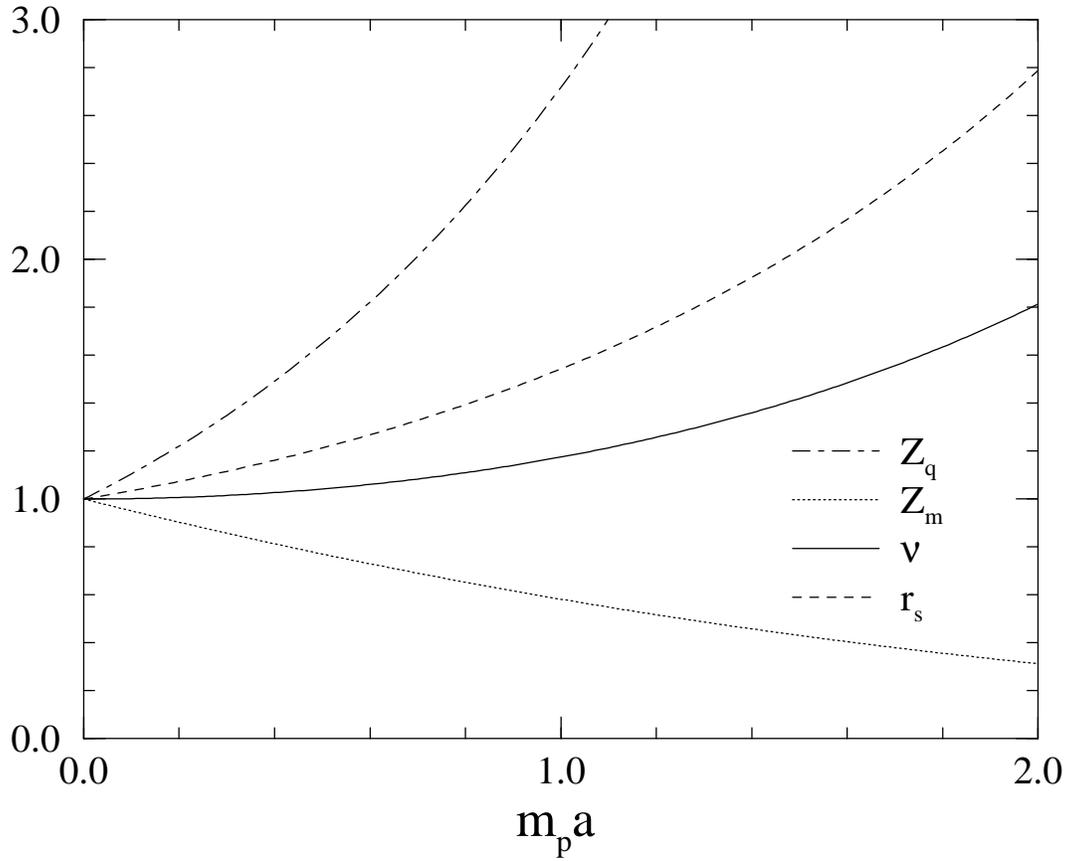,width=140mm,angle=-90}
\vskip +3mm 
}
\caption{Tree-level values for $Z_q$, $Z_m$, $\nu$ and
$r_s$ as functions of $m_p a$. We choose $r_t=1$.} 
\label{fig:tree_ma}
%\vspace{-8mm}
\end{figure}

$Z_q$, $Z_m$, $\nu$ and $r_s$ can be determined by
demanding  that the tree-level quark propagator 
$S_q(p)$ derived from eq.~(\ref{eq:qaction_iso}) should reproduce
the relativistic form 
\be
S_q(p_0,p_i)=
\frac{1}{Z_q}
\frac{ -i \gamma_0 p_0- i \sum_i \gamma_i p_i + m_p}
{p_0^2+\sum_i p_i^2 +m_p^2}
+{\rm (no\;\; pole\;\; terms)}+O((p_i a)^2)
\label{eq:qprop_r_iso}
\ee
around the pole.
Imposing $p_i=0$ we first obtain 
\ben
m_p&=&{\rm log}\left|\frac{m_0+r_t+\sqrt{m_0^2+2r_t m_0+1}}
{1+r_t}\right|,\\
Z_m&=&\frac{m_p}{m_0},\\
Z_q&=&{\rm cosh}(m_p)+r_t {\rm sinh}(m_p).
\een
We then find with finite spatial momenta
\ben
\nu&=&\frac{{\rm sinh}(m_p)}{m_p},
\label{eq:nu}\\
r_s&=&\frac{{\rm cosh}(m_p)+r_t {\rm sinh}(m_p)}{m_p}
-\frac{{\rm sinh}(m_p)}{m_p^2}.
\label{eq:rs}
\een
We should notice that the $m_Q a$ corrections start at 
$O((m_Q a)^2)$, not $O(m_Q a)$, in the $\nu$ parameter as expected.
Figure~\ref{fig:tree_ma} illustrates the $m_p a$ dependences of
$Z_q$, $Z_m$, $\nu$ and $r_s$ in case of $r_t=1$. We observe that
the $m_p a$ dependences of $\nu$ is relatively
mild compared to those of $Z_q$, $Z_m$ and $r_s$. 

\begin{figure}[h]
\centering{
\hskip -0.0cm
\psfig{file=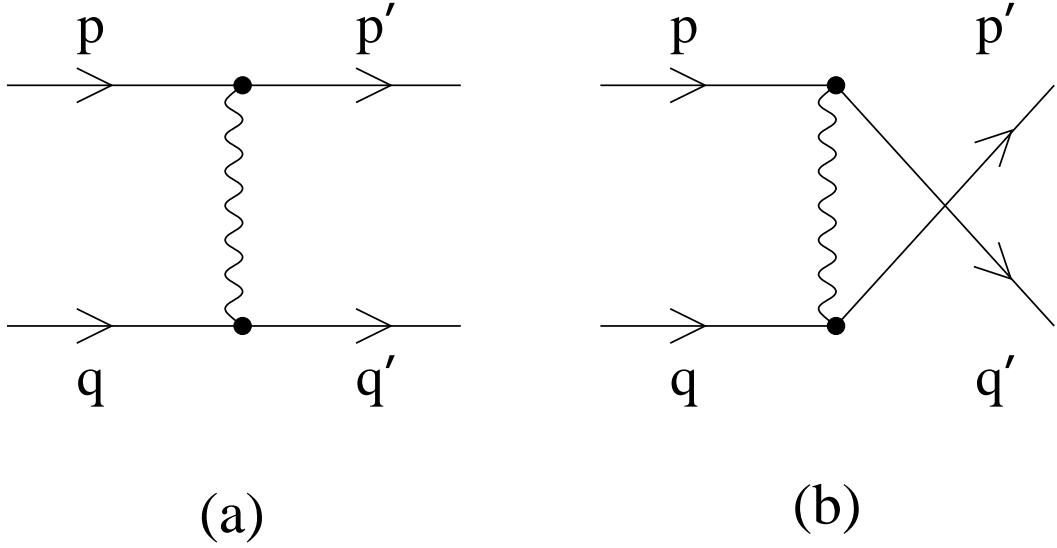,width=140mm,angle=-90}
\vskip +3mm 
}
\caption{Tree-level diagrams for the quark-quark scattering.} 
\label{fig:scatt}
%\vspace{-8mm}
\end{figure}

To fix the $c_E$ and $c_B$ parameters we
consider the quark-quark scattering amplitude 
depicted in Fig.~\ref{fig:scatt}.
The improvement condition is that
$c_E$ and $c_B$ should
be chosen to reproduce the following form of
scattering amplitude at the on-shell point 
removing the $m_Q a$ corrections,
\ben
T &=&-g^2{\bar u}(p^\prime)\gamma_\mu u(p)D_{\mu\nu}(p-p^\prime)
{\bar u}(q^\prime)\gamma_\nu u(q)\nn\\
&&-g^2{\bar u}(q^\prime)\gamma_\mu u(p)D_{\mu\nu}(p-q^\prime)
{\bar u}(p^\prime)\gamma_\nu u(q)
+O((p_i a)^2,(q_i a)^2,(p^\prime_i a)^2,(q_i^\prime a)^2),
\een
where $D_{\mu\nu}$ is the gluon propagator on the lattice.
Notice that with the use of the Gordon identity
(; on-shell condition for external spinors $u$, ${\bar u}$)
the quark-gluon interaction induced by the
clover term can be transformed into the ordinary quark-gluon
vertex:
\ben
&&{\bar u}(p^\prime)\sum_l i \sigma_{0l}
({\rm sin}(p_l^\prime)-{\rm sin}(p_l)) u(p) \nn\\
&=&{\bar u}(p^\prime)\frac{1}{\nu}[
i{\rm sin}(p_0^\prime)+i{\rm sin}(p_0)
+\gamma_0\{ 2(m_0+r_t)-r_t({\rm cos}(p^\prime_0)+{\rm
cos}(p_0))\nn\\&&
+r_s\sum_l (2-{\rm cos}(p^\prime_l)-{\rm cos}(p_l))\}] u(p),
\label{eq:gordon_t}\\
&&{\bar u}(p^\prime) i \sigma_{i0}
({\rm sin}(p_0^\prime)-{\rm sin}(p_0)) u(p) 
+{\bar u}(p^\prime)\sum_{l}\nu i \sigma_{il}
({\rm sin}(p_l^\prime)-{\rm sin}(p_l)) u(p) \nn\\
&=&{\bar u}(p^\prime)[{\nu}
(i{\rm sin}(p_i^\prime)+i{\rm sin}(p_i))
+\gamma_i\{ 2(m_0+r_t)-r_t({\rm cos}(p^\prime_0)+{\rm cos}(p_0))\nn\\&&
+r_s\sum_l (2-{\rm cos}(p^\prime_l)-{\rm cos}(p_l))\}] u(p).
\label{eq:gordon_s}
\een
This improvement procedure with the finite quark mass is 
an extension of the previous work\cite{c_sw} that
determined the $c_{\rm SW}=c_E=c_B$ parameter up to
one-loop level in the massless case.
After some algebra with the aid of eqs.~(\ref{eq:gordon_t}) 
and (\ref{eq:gordon_s}), we obtain
\ben
c_E&=&r_t \nu,
\label{eq:ce}\\
c_B&=&r_s,
\label{eq:cb}
\een
where $\nu$ and $r_s$ are already determined from the
on-shell improvement on the quark propagator.

You may have already noticed that our values for $\nu$, $r_s$ and
$c_E$ are different from those derived in Ref.~\cite{fermilab}.
This difference originates from whether the on-shell
improvement is implemented in the relativistic way or in the
nonrelativistic way. (more precisely, in case that the
Lagrangian does not retain the 
rotational invariance on the Euclidean space-time,
we need both the momentum operator and the Hamiltonian
to discuss the rotational symmetry.)
For example, both methods give the same
relation for the $\nu$ and $r_s$ parameters from the
dispersion relation:
\be
\nu^2+r_s{\rm sinh}(m_p)=\frac{{\rm sinh}(m_p)}{m_p}
\left\{{\rm cosh}(m_p)+r_t {\rm sinh}(m_p)\right\}.
\ee
In the nonrelativistic approach, however, $\nu$ and $r_s$
are not distinguishable due to the lack of relativistic
informations.
Generally speaking, we do not have sufficient number of
nonrelativistic conditions
to fix the coefficients of the 
relativistic operators with higher dimensions because the
degrees of freedom of quarks are
smaller  
in the nonrelativistic approximation compared to the
relativistic case.
Although the nonrelativistic approach with the Wilson type
quark action\cite{fermilab} has been considered to work better
than the NRQCD in the charm quark region where
the sizable relativistic effects are expected,
this is not necessarily assured because this approach  
does not meet the relativistic on-shell improvement. 

\subsection{Large $m_Q a$ and static limit}
\label{subsec:static}

Although we have restricted ourselves to the case of finite
$m_Q a$ so far,
it is worthwhile to show that we can 
derive the static quark action from 
eq.~(\ref{eq:qaction_iso}) by taking $m_Q a\rightarrow \infty$.
In terms of the heavy quark field $h(x)$ defined by
\ben
q(x) &=& \frac{{\rm e}^{iE t}}{\sqrt{m_0}} h(x),
\een
where $E = \pm i m_p$,
the lattice action in eq.~(\ref{eq:qaction_iso}) becomes
\ben
 S_q^{\rm imp} &=& \sum_x \left[
\left({1+ \frac{r_t}{m_0}+\frac{3r_s}{m_0}}\right) \bar h(x) h(x) -
\frac{{\rm e}^{iE}}{2m_0}\bar h(x) (r_t -\gamma_0) 
U_0(x) h(x+\hat 0)\right. \nn \\
&&\left.-\frac{{\rm e}^{-iE}}{2m_0}\bar 
h(x+\hat 0) (r_t +\gamma_0) U_0(x)^\dagger  h(x)\right] 
+ O\left(\frac{(\nu, c_E, r_s, c_B)}{m_0}\right) .
\een
Taking $m_0\rightarrow \infty$ and setting $r_t=1$ for simplicity, 
we obtain
\ben
S_q^{\rm imp} &\simeq& \sum_x \left[\bar h(x) h(x) -
\bar h(x+\hat 0) \frac{1 +\gamma_0}{2} U_0(x)^\dagger  h(x) \right]
\een
for the heavy quark($E = i m_p$), or
\ben
S_q^{\rm imp} &\simeq& \sum_x \left[\bar h(x) h(x) -
\bar h(x) \frac{1 -\gamma_0}{2} U_0(x) h(x+\hat 0) \right]
\een
for the heavy anti-quark($E = - i m_p$),
since
\ben
\frac{\nu}{m_0}=\frac{\ce}{m_0}&\sim& O\left(\frac{1}{m_p}\right),
\label{eq:nu/m0}\\
\frac{r_s}{m_0}=\frac{\cb}{m_0}&\sim& O\left(\frac{1}{m_p}\right),
\label{eq:rs/m0} \\
\frac{{\rm e}^{m_p}}{m_0} &\simeq& 1,
\een
where $m_p={\rm log}|1+m_0|$ is used.
We replace $\frac{1\pm\gamma_0}{2} h(x) = h(x)$
from the property that $ \gamma_0 h(x) = h(x)$ for the quark and
$\gamma_0 h(x) = -h(x)$ for the anti-quark.
Thus we exactly obtain the static quark action, where the quark moves forward,
or the static anti-quark action, where the anti-quark moves backward
in time.

The existence of the static limit may give some constraints on the
mass dependent scaling violation for $m_Q a \gg 1$.
As an explicit example, we take the heavy-light pseudoscalar
meson decay constant $f_{HL}$.
One can extract $f_{HL}$ from the correlation function of the
heavy-light current 
${\cal A}^{HL}_0(x) = \bar q_H(x)\gamma_0\gamma_5 q_L(x)$
with zero spatial momentum following
\ben
Z_q^{(0)}\langle {\cal A}^{HL}_0 (t) {\cal A}^{HL}_0(0)\rangle
&\simeq& f_{HL}^2 m_{HL} e^{-m_{HL} t},
\een
where $Z_q^{(0)}$ denotes the renormalization factor of the 
quark wave function at the tree-level. 
On the other hand, in the static limit, the current is given by
$ \bar h(x) \gamma_0 \gamma_5 q_L(x) =
\sqrt{m_0}{{\rm e}^{m_Qt}}{\cal A}^{HL}_0(x)$, and
the correlation function
\ben
m_0 \langle {\cal A}^{HL}_0 (t) {\cal A}^{HL}_0(0)\rangle &\simeq &
C^2 e^{-\Delta E t}
\een
has a well-defined limit. 
 From the relation $ Z_q^{(0)}={\rm e}^{m_p}\simeq m_0$ with
$r_t=1$ and
$ m_{HL} \simeq m_Q + \Delta E$, we obtain
\ben
f_{HL} &=& \frac{ C}{ \sqrt{m_{HL}}} .
\een
Since the continuum limit can be taken in the static theory,
$ C$ should behave as 
\ben
\frac{C}{\lqcd^{3/2}} &=& w_0 + w_k (a\lqcd)^k +O( (a\lqcd)^{k+1})
\een
where $k=1$ for the Wilson light quark action or 
$k=2$ for the (nonperturbatively tuned) SW light quark action. 
Therefore the following relation holds 
for $m_Q a\rightarrow \infty$:
\ben
\frac{f_{HL}}{\lqcd} &=& \sqrt{\frac{\lqcd}{m_{HL}}} 
\left[ w_0 + w_k (a\lqcd)^k +O( (a\lqcd)^{k+1})\right].
\label{eq:fhl_sta}
\een
On the other hand, the decay constant should behave
\ben
\frac{f_{HL}}{\lqcd} &=& v_0(\lqcd/m_Q)\nn\\
&&\times\left[1 + v_k^L ( a\lqcd )^k  
+ v_n^H (m_Q a) ( a\lqcd )^n + 
O( (a\lqcd )^{k+1},(a\lqcd )^{n+1})\right]
\label{eq:fhl_rel}
\een
from our consideration in the previous sections, where 
$v_k^L$ is the scaling violation caused by the light quark action,
and $v_n^H(m_Q a)$ comes mainly from the Wilson/SW heavy quark action
for $ n=1$ or from the action of eq.~(\ref{eq:qaction_iso}) with 
the nonperturbatively tuned $\nu$, $r_s$, $c_E$ and $c_B$ for
$n=2$.
Comparing eqs.~(\ref{eq:fhl_sta}) and (\ref{eq:fhl_rel}), we find
\ben
v_0(\lqcd/m_Q) &\rightarrow& \sqrt{\frac{\lqcd}{m_{HL}}}  w_0 
\een
and
\ben
v_k^L &\rightarrow&  w_k, \qquad v_n^H (m_Q a)  \rightarrow 0 
\qquad \mbox{for $ n < k$} \\
v_k^L+v_n^H (m_Q a) &\rightarrow&  w_k, \qquad \mbox{for $ n = k$} \\
v_k^L &\rightarrow&  w_k, \qquad v_n^H (m_Q a)  \rightarrow w_n 
\qquad \mbox{for $ n > k$} 
\een
in the limit $m_Q a \rightarrow\infty$,
where $w_k$ and $w_n$ are universal constants independent of the choice of
the heavy quark action.
Note that we expect $v_0$ can be expanded 
in terms of of $m_q/\lqcd$ near the chiral limit. 

The important point is that the function $v_n^H ( m_Q a)$ 
becomes constant (or even vanishes)
in the limit $m_Q a \rightarrow\infty$. On the other hand,
we also know that the behavior of $v_n^H ( m_Q a)$ is
benign under the condition $m_Q a\simlt 1$. 
 From these facts it would be reasonable to assume that
the function $v_n^H ( m_Q a)$ behaves modestly all over the
range of $m_Q a$ from the
massless limit to the static limit.
This assumption is supported by previous numerical
results for the heavy-light decay constants $f_{D_s}$ and $f_{B_s}$
which are obtained by using the pole mass under the
condition $m_Q a\simgt 1$
with the Wilson quark action. 
Figures 4 and 5 in
Ref.\cite{jlqcd96} show that the naive continuum extrapolation with
a linear form for
$f_{D_s}$ and $f_{B_s}$
gives ``reasonable'' values in the 
continuum limit.
This outcome is understood as follows:
Since $v_n^H ( m_Q a)$ is a modest function in terms of
$m_Q a$, the leading scaling violation effects
$v_1^H ( m_Q a)a\lqcd$ for $f_{D_s}$ and $f_{B_s}$ 
are effectively removed by the naive linear extrapolation.
However, the $O(a\lqcd)$ cutoff
effects cannot be completely removed. The remnant is
estimated to be 
$|v_1^H (m_Q a_{\rm max})-v_1^H (m_Q a_{\rm min})|a_{\rm av}\lqcd$, where 
$a_{\rm max}$ and $a_{\rm min}$ are the maximum
and minimum of the lattice spacing used for the continuum  
extrapolation, 
and $a_{\rm av} = a_{\rm max} a_{\rm min}/
(a_{\rm max}-a_{\rm min})$.

Although we have focused on the case of heavy-light
pseudoscalar meson decay constants,
the above arguments on scaling violation effects 
can be easily generalized to any observables which
can be defined in the static limit.

\subsection{Nonperturbative renormalization}
\label{subsec:npr}

Let us turn to a nonperturbative determination of 
$Z_q$, $Z_m$, $\nu$, $r_s$, $\cb$ and $\ce$.
We first consider the $\nu$ and $r_s$ parameters.
Since the rotational symmetry breaking due to the
$m_Q a$ corrections  deviates the $\nu$ and $r_s$ parameters
from one, it would be a reasonable way to adjust them such that
the correct dispersion relations are reproduced for some
hadronic states. We think  
a set of dispersion relations for the
heavy-heavy and heavy-light mesons is a good choice. 
A previous study demonstrated a clear distinction
between the the kinetic
masses defined by $\partial^2 E_{\rm HH}/\partial p_i^2$
for the heavy-heavy meson and
$\partial^2 E_{\rm HL}/\partial p_i^2$ for the heavy-light meson
(see Fig.~1 in Ref.~\cite{jlqcd96}), 
which tells us that the dispersion relations for the
heavy-heavy and heavy-light mesons give two independent
conditions to fix both $\nu$ and $r_s$ parameters.
We point out that to avoid the ambiguities 
coming from the clover term
it would be better to consider the dispersion relations of
the spin averaged meson states over the pseudoscalar and 
vector channels. This is motivated by an 
observation that the clover term causes the hyperfine splitting.

Nonperturbative determination of $\cb$ and $\ce$ is a little bit troublesome.
Although in the massless case the clover coefficient 
can be determined with the
aid of the PCAC relation,
this method does not work with the massive case. 
The reason is that
the chiral symmetry allows 
the clover terms with the odd power of $ma$ corrections
$(ma)^{2n+1}i{\bar q}(x)\sigma_{\mu\nu}F_{\mu\nu} q(x)$
$(n\ge 0)$
as discussed in Sec.~\ref{subsec:chiral}.  
This implies that even the chirally symmetric quark actions,
{\it e.g.,} the domain wall and the overlap quark actions,
suffer from this difficulty. 
However, we can at least evaluate
$\cb$ and $\ce$ perturbatively up to one-loop level 
by extending the calculation in Sec.~\ref{subsec:tree}.
In this case the remaining
cutoff effects are of order $\alpha^2 a\lqcd$, which might be
small enough for numerical studies.

\begin{figure}[h]
\centering{
\hskip -0.0cm
\psfig{file=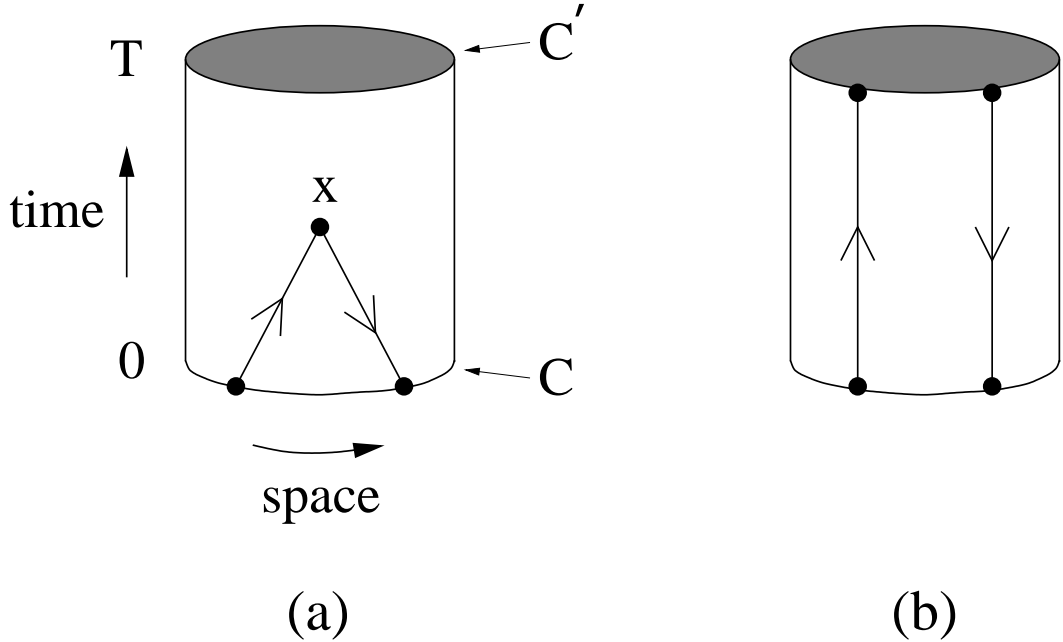,width=140mm,angle=-90}
\vskip +3mm 
}
\caption{Quark diagrams contributing to (a) $f_{A,P}(x)$ and
(b) $f_{1}$. ${\cal O}_{A,P}$ are inserted at the point
$x$. $C$ and $C^\prime$ denote the boundary conditions for
the gauge fields at $t=0$ and $T$.} 
\label{fig:sf}
%\vspace{-8mm}
\end{figure}

As for a nonperturbative determination of 
renormalization factors, 
we consider the use of the Schr\"{o}dinger 
functional(SF) method\cite{sf}.
The renormalization of the quark mass is made through the
renormalizations of the axial current and the pseudoscalar
density using the PCAC relation:
\be
{\bar m}(\mu=1/L)=
\frac{Z_A({\bar g},{\bar m}L,g,ma)
\partial_\mu {\bar q}(x)\gamma_\mu\gamma_5 q(x)}
{Z_P({\bar g},L,{\bar m}L,g,ma){\bar q}(x)\gamma_5 q(x)}.
\label{eq:pcac}
\ee
where $L$ is the physical box size and 
${\bar g}$ and ${\bar m}$ are the renormalized coupling and
quark mass in the SF scheme.
$Z_A$ and $Z_P$ with the finite quark mass are defined by
\ben
Z_A({\bar g},{\bar m}L,g,ma)&=&
\frac{\sqrt{c_m f_{1}}}{f_A(x_0=L/2)},\\
Z_P({\bar g},L,{\bar m}L,g,ma)&=&
\frac{\sqrt{c_m f_{1}}}{f_P(x_0=L/2)},
\een
where $c_m$ is a mass-dependent constant, and
$f_A$,  
$f_P$ and $f_{1}$ are
the correlation functions given by
\ben
f_A(x_0)&=&-\frac{1}{3}\int d^3{\vec y}d^3{\vec z}
\la {\bar q}(x)\gamma_\mu\gamma_5 q(x)
{\bar \zeta}({\vec y})\gamma_5 \zeta({\vec z}) \ra ,\\
f_P(x_0)&=&-\frac{1}{3}\int d^3{\vec y}d^3{\vec z}
\la {\bar q}(x)\gamma_5 q(x)
{\bar \zeta}({\vec y})\gamma_5 \zeta({\vec z}) \ra ,\\
f_{1}&=&-\frac{1}{3L^6}\int d^3{\vec u}d^3{\vec v} d^3{\vec y}d^3{\vec z}
\la {\bar \zeta}^\prime({\vec u})\gamma_5 \zeta^\prime({\vec v})
{\bar \zeta}({\vec y})\gamma_5 \zeta({\vec z}) \ra
\een
with $\zeta$, $\zeta^\prime$, ${\bar \zeta}$ and 
${\bar \zeta}^\prime$ the boundary quark fields. 
We
illustrate the function $f_{A,P}$ and $f_{1}$ in Fig.~\ref{fig:sf}.
As for the boundary conditions we refer to the description
in Ref.~\cite{qmass}.

Although each of $Z_A$, $Z_P$, 
$\la {\bar q}(x)\gamma_\mu\gamma_5 q(x)\ra$ and 
$\la {\bar q}(x)\gamma_5 q(x)\ra$ has the power corrections
of $ma$, they are canceled out in the combinations 
$Z_A\la {\bar q}(x)\gamma_\mu\gamma_5 q(x)\ra$ and
$Z_P\la {\bar q}(x)\gamma_5 q(x)\ra$ and
the $O((a\Lambda_{\rm QCD})^2)$ uncertainties are 
left. 
This assures us to take the continuum limit for
the renormalized matrix elements and also for the
renormalized quark mass of eq.~(\ref{eq:pcac}).
These quantities, however, are
defined in the SF scheme with the finite quark mass
and hence different from those
renormalized in the SF scheme at
the massless point, which are exactly what we want.
Therefore, we still need the finite renormalization factor
to make the conversion between the two renormalization descriptions.
This can be obtained by taking the continuum limit of the ratio
\ben
\frac{Z_A({\bar g},{\bar m}L=0,g^\prime,ma^\prime=0)}
{Z_A({\bar g},{\bar m}L,g^\prime,ma^\prime)},\\
\frac{Z_P({\bar g},L,{\bar m}L=0,g^\prime,ma^\prime=0)}
{Z_P({\bar g},L,{\bar m}L,g^\prime,ma^\prime)},
\een 
for several $g^\prime$ chosen to satisfy $ma^\prime\ll 1$.
We remark that the physical size $L$ for the SF scheme can
be taken to be much smaller than that for the measurement of
the spectral quantities and the various matrix elements,
allowing us to access the finer lattice spacing with $g^\prime$.

\section{Conclusions}
\label{sec:conclusion}

In this paper we have first examined the validity 
of the idea of anisotropic lattice by making a perturbative
calculation of the quark self energy up to $O(g^2 a)$. 
Our results show that the $m_Q a_s$ corrections revive 
through one-loop diagram. 
We also find that on the anisotropic lattice
the four parameters must be adjusted to
remove all the terms of order $a$ even in the massless case.  
 From a theoretical point of view
the anisotropic lattice is not necessarily advantageous 
over the isotropic one.

In the second part of this paper we have presented a new
relativistic approach to the heavy quarks on the lattice.
The idea is based on the relativistic on-shell improvement with
the finite $m_Q a$ corrections.
We have shown that the cutoff effects can be reduced to
$O((a\Lambda_{\rm QCD})^2)$ putting
the $(m_Q a)^n$ corrections
on the renormalization factors of the quark mass $Z_m$ and
wave function $Z_q$. 
As far as the spectral quantities such as hadron masses are concerned,
the $(m_Q a)^n$ corrections in $Z_m$ and $Z_q$
do not matter:
$Z_q$ does not affects on the spectral
quantities and the $(m_Q a)^n$ corrections 
in $Z_m$ can be handled by employing the pole
mass fixed from some spectral quantity. 
On the other hand, in case of calculating
the quark mass or the various hadron matrix elements
we can control them by determining
the renormalization factors nonperturbatively.

The leading scaling violation for 
various types of actions are summarized as follows.
$f_1(m_Q a) a\lqcd $ for the Wilson quark
($f_1(0)\not=0$) and the $O(a)$ improved SW quark ($f_1(0)=0$), while
$f_2(m_Q a) (a\lqcd)^2$ for our proposed action with mass-dependently
tuned $\nu$, $r_s$, $c_E$ and $c_B$. Therefore, 
if the magnitude of $f_1( m_Q a)$ is $O(1)$,
the scaling violation for heavy hadron masses
with the $O(a)$ improved SW quark might 
be as bad as that for light hadron masses
with the ordinary Wilson quark.
For sufficiently small $m_Q a$,
we can remove the leading scaling violations by
extrapolating the data at several lattice spacings 
to the continuum limit.
Even if $m_Q a\sim O(1)$, 
$f_2(m_Q a) (a\lqcd)^2$ in our proposed action is expected  
to be negligibly small in case of $(a\lqcd)^2\sim 0.01$

Our relativistic approach has the strong point over the
nonrelativistic ones:  
the finer the lattice spacing becomes, the better the 
approach works. This is a desirable feature because
we can take the full advantage of configurations with
finer lattice spacing generated to control the cutoff
effects on the light hadron physics.
We are going to perform a numerical simulation to test the
ideas presented in this paper.

\acknowledgements

One of us (Y.K.) thanks M.~Okawa for useful discussions.
This work is supported in part by the Grants-in-Aid of the Ministry of 
Education (Nos. 13740169,12014202,12640253). 

%\section*{appendix}

%\input{ref.tex}

%\end{document}

%%%%%%%%%%%%%%%%%% ref.tex %%%%%%%%%%%%%%%%

\newpage

\newpage

\newpage

\newpage

\newpage

\newpage

%\newpage
\vspace{10mm}

\newpage

\begin{table}[h]
\begin{center}
\caption{\label{tab:action}Expected magnitude of the cutoff
effects due to the higher
dimensional operators in the quark action. 
%$m_q$ and 
$m_Q$ denote the %light and
heavy quark masses. }
\begin{tabular}{llll}
\multicolumn{2}{c}{operator} & light & heavy  \\
\hline
${\cal O}_{5b}:$ & $i\sum_i {\bar q}(x)\sigma_{0i}F_{0i} q(x)$
                 & $a \Lambda_{\rm QCD}$ 
                 & $a \Lambda_{\rm QCD}^2/m_Q$ \\
                 & $i\sum_{ij} {\bar q}(x)\sigma_{ij}F_{ij} q(x)$
                 & $a \Lambda_{\rm QCD}$ 
                 & $a \Lambda_{\rm QCD}$ \\
%${\cal O}_{6d}:$ & $mi\sum_i {\bar q}(x)\sigma_{0i}F_{0i} q(x)$
%                 & $m_q a^2 \Lambda_{\rm QCD}$
%                 & $(a \Lambda_{\rm QCD})^2$ \\
%                 & $mi\sum_{ij} {\bar q}(x)\sigma_{ij}F_{ij} q(x)$
%                 & $m_q a^2 \Lambda_{\rm QCD}$
%                 & $m_Q a^2 \Lambda_{\rm QCD}$ \\
${\cal O}_{6a}:$ & ${\bar q}(x)\gamma_0 D_0^3 q(x)$ 
                 & $(a \Lambda_{\rm QCD})^2$
                 & 
                   $(m_Q a)^3 {\bar q}(x) q(x)$,
                   $(m_Q a)^2 {\bar q}(x)\gamma_i D_i q(x)$,
                   $(m_Q a) {\bar q}(x)D_i^2 q(x)$ \\ 
                 & ${\bar q}(x)\gamma_i D_i^3 q(x)$
                 & $(a \Lambda_{\rm QCD})^2$ 
                 & $(a \Lambda_{\rm QCD})^2$\\
${\cal O}_{6d}:$ & $i\sum_i {\bar q}(x)\gamma_i[D_0,F_{i0}] q(x)$
                 & $(a \Lambda_{\rm QCD})^2$  
                 & $a^2 \Lambda_{\rm QCD}^3/m_Q$ \\
                 & $i\sum_i {\bar q}(x)\gamma_0[D_i,F_{0i}]q(x)$
                 & $(a \Lambda_{\rm QCD})^2$ 
                 & $(a \Lambda_{\rm QCD})^2$ \\
                 & $i\sum_i {\bar q}(x)\gamma_i[D_j,F_{ij}]q(x)$
                 & $(a \Lambda_{\rm QCD})^2$
                 & $a^2 \Lambda_{\rm QCD}^3/m_Q$ \\
${\cal O}_{6f}:$ &
                 & $(a \Lambda_{\rm QCD})^2$ 
                 & $(a \Lambda_{\rm QCD})^2$ \\
%${\cal O}_{7 }:$ & ${\bar q}(x)D_0^4 q(x)$ 
%                 & $(a \Lambda_{\rm QCD})^3$
%                 & 
%                   $(m_Q a)^4 {\bar q}(x) q(x)$,
%                   $(m_Q a)^2 {\bar q}(x)D_i^2 q(x)$ \\ 
%                 & ${\bar q}(x)D_i^4 q(x)$
%                 & $(a \Lambda_{\rm QCD})^3$
%                 & $(a \Lambda_{\rm QCD})^3$
\end{tabular}
\end{center}
\end{table}

%\begin{table}[h]
%\begin{center}
%\caption{\label{tab:axial}Expected magnitude of the cutoff
%effects due to the higher
%dimensional operators in the axial current. 
%$m_q$ and $m_Q$ denote the light and
%heavy quark masses. }
%\begin{tabular}{lll}
%operator & light-light & heavy-light  \\
%\hline
%$({\cal A}_{4a})_0$ & $a \Lambda_{\rm QCD}$ 
%                    & $a \Lambda_{\rm QCD}$ \\
%$({\cal A}_{4a})_i$ & $a \Lambda_{\rm QCD}$ 
%                    & $a \Lambda_{\rm QCD}$ \\
%$({\cal A}_{5a})_0$ & $(a \Lambda_{\rm QCD})^2$
%                    & $(a \Lambda_{\rm QCD})^2$ \\
%$({\cal A}_{5a})_i$ & $(a \Lambda_{\rm QCD})^2$
%                    & $(a \Lambda_{\rm QCD})^2$ \\
%$({\cal A}_{5e})_0$ & $(a \Lambda_{\rm QCD})^2$
%                    & $(a \Lambda_{\rm QCD})^2$ \\
%$({\cal A}_{5e})_i$ & $(a \Lambda_{\rm QCD})^2$
%                    & $(a \Lambda_{\rm QCD})^2$ \\
%\end{tabular}
%\end{center}
%\end{table}

\end{document}